\documentclass{emulateapj}
\shorttitle{SMA $^{12}$CO(J = 6 -- 5) and $435~\mu$m observations of Arp 220}
\shortauthors{Matsushita et al.}

\slugcomment{Accepted for publication in The Astrophysical Journal}
\begin{document}

\title{SMA $^{12}$CO(J = 6 -- 5) and $435~\mu$m interferometric
	imaging of the nuclear region of Arp 220}

\author{Satoki Matsushita\altaffilmark{1},
	Daisuke Iono\altaffilmark{2},
	Glen R. Petitpas\altaffilmark{3},
	Richard C.-Y. Chou\altaffilmark{1,4},
	Mark A. Gurwell\altaffilmark{3},
	Todd R. Hunter\altaffilmark{3,5},
	Jeremy Lim\altaffilmark{1}
	Sebastien Muller\altaffilmark{1},
	Alison B. Peck\altaffilmark{3,6},
	Kazushi Sakamoto\altaffilmark{1},
	Satoko Sawada-Satoh\altaffilmark{1,7},
	Martina C. Wiedner\altaffilmark{8},
	David J. Wilner\altaffilmark{3},
	Christine D. Wilson\altaffilmark{9}
	}

\altaffiltext{1}{Academia Sinica Institute of Astronomy and
	Astrophysics, P.O.\ Box 23-141, Taipei 10617, Taiwan, R.O.C.}
\altaffiltext{2}{National Astronomical Observatory of Japan,
	2-21-1 Osawa, Mitaka, Tokyo 181-0015, Japan}
\altaffiltext{3}{Harvard-Smithsonian Center for Astrophysics,
	60 Garden St., MS-78, Cambridge, MA 02138}
\altaffiltext{4}{Department of Astronomy and Astrophysics,
	University of Toronto, 50 St.\ George Street, Toronto,
	ON M5S 3H4, Canada}
\altaffiltext{5}{National Radio Astronomy Observatory,
	520 Edgemont Road, Charlottesville, VA 22903-2475}
\altaffiltext{6}{Joint ALMA Office, Av El Golf 40, Piso 18, Santiago,
	Chile}
\altaffiltext{7}{Department of Physics, Faculty of Science,
	Yamaguchi Univerity, 1677-1 Yoshida, Yamaguchi, Yamaguchi,
	752-8512 Japan}
\altaffiltext{8}{1.\ Physikalisches Institut, Universit\"at zu K\"oln,
	50937 K\"oln, Germany}
\altaffiltext{9}{Department of Physics and Astronomy,
	McMaster University, Hamilton, ON L8S 4M1, Canada}

\begin{abstract}
We have used the Submillimeter Array (SMA) to make the first
interferometric observations (beam size $\sim1\arcsec$, or
$\sim400$~pc) of the $^{12}$CO J=6-5 line and 435~$\mu$m (690~GHz)
continuum emission toward the central region (half power field of
view $17\arcsec$) of the nearby ultra-luminous infrared galaxy
(ULIRG) Arp 220.
These observations resolve the eastern and western nuclei from each
other, in both the molecular line and dust continuum emission.
At 435~$\mu$m, the peak intensity of the western nucleus is stronger
than the eastern nucleus, and the difference in peak intensities is
less than at longer wavelengths.
Fitting a simple model to the dust emission observed between 1.3~mm
and 435~$\mu$m suggests that dust emissivity power law index in the
western nucleus is near unity and steeper in the eastern nucleus,
about 2, and that the dust emission is optically thick at the shorter
wavelength.
Comparison with single dish measurements indicate that the
interferometer observations are missing $\sim60\%$ of the dust
emission, most likely from a spatially extended component to which
these observations are not sensitive.
The $^{12}$CO J=6-5 line observations clearly resolve kinematically
the two nuclei.
The distribution and kinematics of the $^{12}$CO J=6-5 line appear to
be very similar to lower J CO lies observed at similar resolution.
Analysis of multiple $^{12}$CO line intensities indicates that the
molecular gas in both nuclei have similar excitation conditions,
although the western nucleus is warmer and denser.
The excitation conditions are similar to those found in other extreme
environments, including the nearby starburst galaxy M82, the AGN
hosting ULIRG Mrk 231, and the high-z QSO BR 1202--0725.
Simultaneous lower resolution observations of the $^{12}$CO,
$^{13}$CO, and C$^{18}$O J=2-1 lines show that the $^{13}$CO and
C$^{18}$O lines have similar intensities, which suggests that both of
these lines are optically thick, or possibly that extreme high mass
star formation has produced in an overabundance of C$^{18}$O.
\end{abstract}


\section{INTRODUCTION}
\label{intro}

Galaxy-galaxy mergers are the phenomena violent enough to disturb the
galactic potentials of both of the colliding galaxies.
They create shocked gas at the merging region, make interstellar
matter fall into the new galactic potential, and induce strong
starbursts.
This extreme starburst heats up the surrounding dust, which sometimes
radiates $10^{12}$ L$_{\odot}$ or more in infrared luminosity.
These sources are often called ultra-luminous infrared galaxies
(ULIRGs).

Since molecular gas and dust create stars, observations of these
components toward ULIRGs are important to understand the nature
of extreme starbursts.
Multiple molecular lines or multiple transition studies revealed that
a large fraction of molecular gas in ULIRGs is dominated by dense and
warm gas \citep[e.g.,][]{sol92,gao04,pap07}.
In addition, the efficiency of star formation is tightly correlated
with the dense gas fraction in molecular gas \citep{sol92,gao04}.

ULIRGs are also important for the study of high-z submillimeter
galaxies.
The Submillimetre Common-User Bolometer Array (SCUBA) on the James
Clerk Maxwell Telescope (JCMT) has detected many high-z submillimeter
galaxies \citep[SMGs; e.g.,][]{hug98,eal99,eal00,sco02,bor03}.
Most of them have infrared luminosities of $\geq10^{12}$ L$_{\odot}$,
and bright ones have $\sim10^{13}$ L$_{\odot}$.
Local ULIRGs can be studied as nearby counterparts of high-z SMGs or
used as nearby templates to derive photometric redshifts.
To compare redshifted emission lines or continuum emission in high-z
galaxies to those in local ULIRGs, we need to observe local ULIRGs at
the same wavelengths as the rest wavelengths of the detected lines or
continuum from the high-z galaxies.
Many of the high-z galaxies were observed at millimeter-wave
\citep[around 1 -- 3~mm; e.g.,][]{gre05,tac06}.
Therefore the rest wavelengths of the detected lines or continuum are
at submillimeter wavelengths.
Our knowledge about submillimeter lines or continuum from local
ULIRGs, especially from compact starburst regions or from nuclei, is
very limited so far, because only recently have high resolution
submillimeter observations been possible.

Here we present the first interferometric $^{12}$CO(6-5) line and
435~$\mu$m (690~GHz) continuum observations of Arp 220 using the
Submillimeter Array \citep[SMA;][]{ho04}.
The interferometric $^{12}$CO(2-1), $^{13}$CO(2-1), and
C$^{18}$O(2-1) lines and 1.3~mm (226~GHz) continuum observations have
also been made simultaneously with the $^{12}$CO(6-5) and
435~$\mu$m observations.
Both $^{12}$CO(6-5) and C$^{18}$O(2-1) lines have never been observed
(even in a single-dish telescope) toward this galaxy so far.
Arp 220 is the nearest ULIRG
\citep[79.9 Mpc, $1\arcsec=387$~pc;][]{san03}, and has therefore been
well studied in various wavelengths.
Optical images show disturbed faint structures \citep[``galaxies
with adjacent loops'';][]{arp66}, which seem like remnants of tidal
tails produced by a galaxy-galaxy merger \citep{jos85}, similar to
numerical simulation results \citep[e.g.,][]{her92,her93}.
The infrared luminosity at $8-1000~\mu$m of this galaxy is
$1.6\times10^{12}$ L$_{\odot}$ \citep{san03},
and high spatial resolution radio continuum \citep{nor88},
near-infrared \citep{gra90}, and mid-infrared \citep{soi99}
observations revealed two nuclei at the center with a separation of
$\sim0.95''$ ($\sim370$ pc).
Molecular gas is very rich in this galaxy,
$\sim9\times10^9$ M$_{\odot}$, and two-thirds of this mass is
concentrated within 400 pc in radius \citep{sco97}.
Such high gas mass concentration is similar also to the numerical
simulation results \citep[e.g.,][]{bar91,mih96}.
These observations and numerical simulations indicate that Arp 220 is
in the final stage of merging.

The past high spatial resolution millimeter-wave interferometric
observations show molecular gas concentrations at the two nuclei, and
these are embedded in an extended ($\sim1$~kpc) molecular structure,
which seems to be a rotating disk coincident with a dust lane in
optical images \citep{sco97,dow98}.
The molecular gas peak at each nucleus shows a steep velocity
gradient with the direction of the gradient different from that of
the large-scale molecular gas disk.
This suggests that a small-scale molecular gas disk rotates around
each nucleus, and both disks are embedded in the large-scale rotating
disk \citep{sak99}.
Recent $\sim0\farcs3$ molecular gas images resolved the detailed
nuclear gas distributions \citep{dow07,sak08}, which are consistent
with the Hubble Space Telescope NICMOS near-infrared imaging results
that suggest an opaque disk around one of the nuclei \citep{sco98}.

Dust emission also peaks at the two nuclei.
\citet{sak99} reported that most of the continuum flux at 1.3~mm
comes from the two nuclei, while \citet{dow98} mentioned that half of
the 1.3~mm continuum flux comes from the extended component.
It is suggested that a few tenths of the 860~$\mu$m \citep{sak08} and
24.5~$\mu$m \citep{soi99} continuum flux comes from the extended
component.
Far-infrared observations with the Infrared Space Observatory (ISO)
also suggest the existence of extended dust emission \citep{gon04}.

\section{OBSERVATIONS AND DATA REDUCTION}
\label{obs}

We observed the center of Arp 220 with the SMA on March 2nd, 2005.
The phase reference center was at
$\alpha(2000) = 15^{\rm h}34^{\rm m}57\fs19$ and
$\delta(2000) = 23\arcdeg30\arcmin11\farcs3$.
The 225~GHz atmospheric opacity was between 0.03 and 0.04, which was
measured at the nearby Caltech Submillimeter Observatory.
Six of the eight 6~m antennas were used with projected antenna
separations between 14~m and 68~m.
High frequency receivers were tuned to observe the redshifted
$^{12}$CO(6-5) line (679.13~GHz) in the lower side band (LSB) and
the 689.13~GHz ($\sim$435~$\mu$m) continuum emission in the upper
side band (USB).
Low frequency receivers were tuned to observe the redshifted
$^{13}$CO(2-1) line (216.47~GHz) and C$^{18}$O(2-1) line (215.64~GHz)
in the LSB and the redshifted $^{12}$CO(2-1) line (226.42~GHz) in
the USB.
The double sideband system temperature for the high frequency band
ranged from 2000~K to 2500~K for most of the time (i.e., at high
elevation), and that for the low frequency band from 140~K to 180~K.
The SMA correlator covers a 2~GHz bandwidth for each sideband of both
high and low frequency bands, which corresponds to the velocity range
of $\sim880$~km~s$^{-1}$ and $\sim2700$~km~s$^{-1}$ for each sideband
of the high and low frequency bands, respectively.
The channel width was configure to have 3.25~MHz
($\sim1.4$~km~s$^{-1}$) and 0.8125~MHz ($\sim1.1$~km~s$^{-1}$) for
the high and low frequency bands, respectively.

We calibrated the data using the Owens Valley Radio Observatory
software package MIR, which is modified for the SMA.
For the high frequency band calibration, we used a partially resolved
source, Callisto, as a gain (amplitude and phase) and flux
calibrator\footnote[9]{The flux values for Callisto and Ceres are
inferred from the SMA Planetary Visibility Function Calculator
(\url{http://sma1.sma.hawaii.edu/planetvis.html})}, since quasars
were too weak for gain calibration.
Callisto was about $50\arcdeg$ away from Arp 220.
The r.m.s.\ phase fluctuation was about $21\arcdeg$.
We used a source model at the gain calibration to correct for the
effect of the partially resolved structure.
Bandpass calibration was done using three sources, Mars, Callisto,
and Ganymede to gain the signal-to-noise ratio (S/N).
Ceres was imaged after the calibrations, and its flux was 22\% lower
than the calculated flux\footnotemark[9].
Although the flux error for Ceres was about 20\%, we conservatively
adopt the flux error of 30\% for the high frequency band data
throughout this paper.
For the low frequency band calibration, we used Ceres, which was
about $40\arcdeg$ away, as the gain calibrator due to its closeness,
and Mars, Callisto, and Ganymede were used as bandpass calibrators.
Callisto was imaged after the calibrations, and showed 21\% lower
than the calculated flux.
Hereafter, we adopt the flux error of 20\% for the low frequency band
data.

\begin{deluxetable*}{cccc}
\tablecaption{Parameters for the continuum and molecular line images
	\label{tab-obs-contline}}
\tablehead{
	\colhead{Wavelength (Frequency)}
		& \colhead{Synthesized Beam Size}
		& \colhead{Velocity Resolution}
		& \colhead{R.M.S. Noise} \\
	\colhead{or Line}
		& \colhead{and Position Angle}
		&
		& \\
	\colhead{[$\mu$m (GHz)]}
		& \colhead{(Linear scale)}
		& \colhead{[km~s$^{-1}$]}
		& \colhead{[mJy~beam$^{-1}$ (mK)]}
	}
\startdata
 435 (689.13)  & $1\farcs2\times0\farcs9$, $139\arcdeg$ &  --- & 190  (450) \\
               & (470~pc $\times$ 350~pc)               &      &            \\
1320 (226.46)  & $3\farcs8\times3\farcs3$, $28\arcdeg$  &  --- & 4.8  (9.1) \\
               & (1.47~kpc $\times$ 1.28~kpc)           &      &            \\
1380 (216.46)  & $3\farcs9\times3\farcs5$, $27\arcdeg$  &  --- & 5.7   (11) \\
               & (1.51~kpc $\times$ 1.36~kpc)           &      & \\ \hline
$^{12}$CO(6-5) & $1\farcs3\times0\farcs8$, $129\arcdeg$ & 30.1 & 535 (1400) \\
               & (500~pc $\times$ 310~pc)               &      &            \\
$^{12}$CO(2-1) & $3\farcs8\times3\farcs3$, $28\arcdeg$  &  5.4 & 32.2  (61) \\
               & (1.47~kpc $\times$ 1.28~kpc)           &      &            \\
$^{13}$CO(2-1), C$^{18}$O(2-1)
               & $3\farcs9\times3\farcs5$, $27\arcdeg$  & 30.8 & 13.1  (25) \\
               & (1.51~kpc $\times$ 1.36~kpc)           &      &
\enddata
\end{deluxetable*}

Data from 5 antennas were used for the $^{12}$CO(6-5) line
imaging because of a correlator problem.
For $435~\mu$m continuum imaging, we used the data from all 6
antennas after discarding the problematic data.
We subtracted the continuum emission from the line emission data
before line imaging.
Since the $^{12}$CO(6-5) line width is comparable with the band
width, we could not obtain the continuum emission from the same
sideband.
We therefore subtracted the continuum emission using the other
sideband (i.e., USB) in the $uv$ plane using the National Radio
Astronomy Observatory software package AIPS.
For the low frequency (1~mm) data, we created the continuum image
using the line free channels in the same band.
The line images for the low frequency data were made after the
subtraction of the continuum emission from the data in the $uv$
plane.

The calibrated data were binned, and the final channel maps have
velocity resolutions of about 30~km~s$^{-1}$ for the $^{12}$CO(6-5),
$^{13}$CO(2-1), and C$^{18}$O(2-1) lines, and about 5~km~s$^{-1}$ for
the $^{12}$CO(2-1) line.
We use the radio definition for the LSR velocity in this paper, which
is $v_{\rm LSR} = c (1 - \nu / \nu_{\rm rest})$.
The C$^{18}$O(2-1) line was observed up to the LSR velocity of
5550~km~s$^{-1}$, covering about 72\% of the total line width, since
this line was at an edge of the bandpass.

We CLEANed the images with natural weighting, and the resulting
synthesized beam sizes were about $1\arcsec$ for 690~GHz band
images, and about $3\arcsec-4\arcsec$ for 230~GHz band images.
The beam sizes and the r.m.s.\ noise levels for all the images
are summarized in Table~\ref{tab-obs-contline}.
The half-power width of the primary beam at 690~GHz and 230~GHz are
$17\arcsec$ (6.6~kpc) and $52\arcsec$ (20.1~kpc), respectively.
These sizes are much larger than the sizes of the line/continuum
emitting regions (at most a few arcseconds), and we did not make any
primary beam correction to our images.

Our $^{12}$CO(6-5) images show a systematic position shift of about
$0\farcs7$ from the peak positions of the double nucleus reported in
the previous observations at longer wavelengths
\citep[e.g.,][]{sak99}.
This offset can be explained by the baseline error, which is about
$0.3\lambda$ or less at 690~GHz.
Furthermore, we used the phase calibrator with about $50\arcdeg$ away
from the source.
Hence we shifted the positions of the $^{12}$CO(6-5) line images
based on the peak positions and kinematics information to be
consistent with the previously published results.
The $435~\mu$m continuum image also shows about $0\farcs1$ position
shift, which is again explained by the baseline error, and therefore
shifted the position with the same manner as the $^{12}$CO(6-5) data.
The low frequency band images did not show any noticeable position
shifts, and therefore we did not shift any images.

\section{RESULTS}
\label{res}

\subsection{$435~\mu$m Continuum Emission}
\label{res-cont435}

\begin{figure}
\plotone{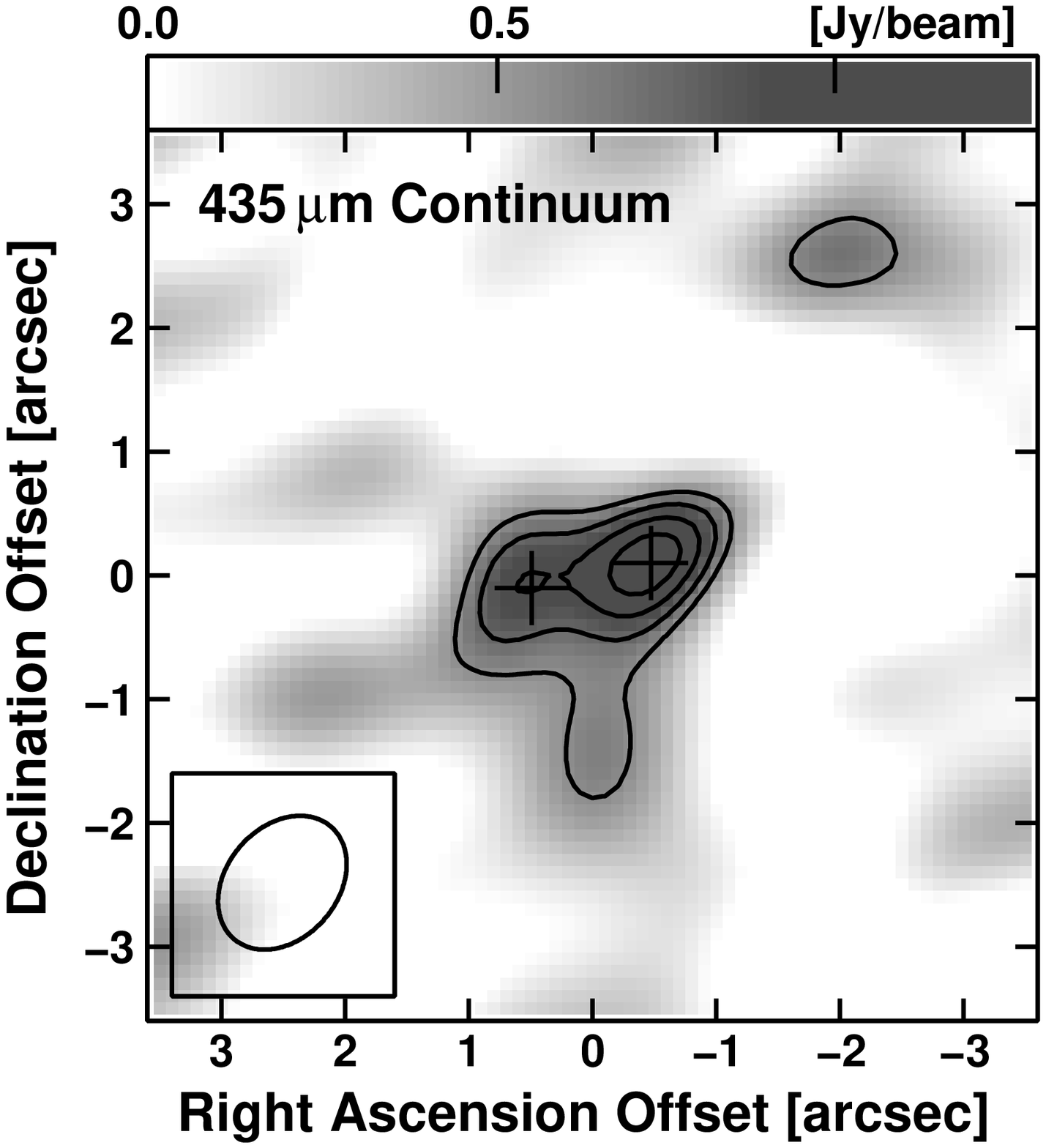}
\caption{
	The $435~\mu$m (689~GHz) continuum emission image of the central
	region of Arp 220.
	The synthesized beam ($1\farcs2\times0\farcs9$ or 470~pc $\times$
	350~pc) with the P.A.\ of $139\arcdeg$ is shown at the
	bottom-left corner.
	The two crosses indicate the 1.3~mm continuum peaks
	\citep{sak99,dow98}, which corresponds to the double nucleus.
	The contour levels are $3, 4, 5,$ and $6\sigma$,
	where $1\sigma$ = 190 mJy beam$^{-1}$.
	The position offsets are measured from
	$\alpha(2000) = 15^{\rm h}34^{\rm m}57\fs25$ and
	$\delta(2000) = 23\arcdeg30\arcmin11\farcs4$.
\label{fig-cont435}}
\end{figure}

The $435~\mu$m (689~GHz) continuum emission image is shown in
Fig.~\ref{fig-cont435}.
The image clearly shows two peaks with a separation of about
$1\arcsec$, consistent with the past continuum images in centimeter,
millimeter, and infrared wavelengths.
We therefore call these peaks eastern and western nuclei as is the
past studies.
Intensities of the western and eastern nuclei are
$1.28\pm0.38$~Jy~beam$^{-1}$ and $0.96\pm0.29$~Jy~beam$^{-1}$,
respectively.
The total flux density of the continuum emission is $2.5\pm0.8$~Jy.
Since the flux distribution can be smeared by the phase fluctuation
or baseline errors, we convolved the image to larger beam sizes (up
to $10\arcsec$ with a Gaussian convolution) and measured the total
flux density.
It did not change from 2.5~Jy, indicating that the flux smearing
effect due to the phase or baseline errors is small.
The r.m.s.\ phase fluctuation at this wavelength was indeed only
$21\arcdeg$ (Sect.~\ref{obs}), which induces $<0\farcs1$
smearing, so it is consistent with this result.

\begin{figure}
\plotone{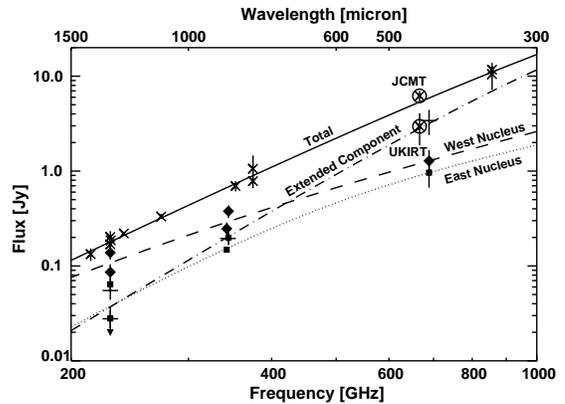}
\caption{Submillimeter wavelength (1.5~mm -- 300~$\mu$m or 200~GHz --
	1000~GHz) SED for various components in Arp 220.
	Crosses, squares, diamonds, and pluses indicate the continuum
	flux density of the entire system, the eastern nucleus, the
	western nucleus, and the extended component, respectively.
	A down-arrow indicates the $3\sigma$ upper limit for the extended
	component.
	Bars on the observational points indicate the observational
	errors.
	Solid, dotted, dashed, and dot-dashed lines are the $\chi^2$
	fitting for each component.
	The low frequency data for the double nucleus and the extended
	component are taken from \citet{dow98} and \citet{sak99} for
	229~GHz, \citet{wie02} for 343~GHz, and \citet{sak08} for
	345~GHz.
	The data for the total flux density are taken from following
	papers:
		214~GHz: \citet{woo89},
		229~GHz: \citet{sco97,dow98,sak99},
		240~GHz: \citet{car92},
		273~GHz: \citet{rig96},
		353~GHz: \citet{dun01},
		375~GHz: \citet{eal89,rig96},
		667~GHz: \citet{eal89,dun01}, and
		857~GHz: \citet{eal89,rig96}.
\label{fig-sed}}
\end{figure}

The previously published $450~\mu$m (670~GHz) single dish results
show a very large variation in the detected flux densities; the
United Kingdom Infrared Telescope (UKIRT) UKT14 result shows the flux
density of $3.0\pm1.1$~Jy \citep{eal89}, but that of the JCMT SCUBA
result shows $6.286\pm0.786$~Jy \citep{dun01}.
We plot the submillimeter (1.5~mm -- $300~\mu$m or
200~GHz -- 1000~GHz) spectrum energy distribution (SED) using
published single dish results in Fig.~\ref{fig-sed}.
The large crosses in the plot are the data for the total flux
densities at various frequencies, and the solid line shows the
$\chi^2$ fitting of the data (see Sect.~\ref{dis-sed} for details).
As shown in the figure, the JCMT SCUBA data point is on the fit, but
the UKIRT UKT14 data point is significantly lower than the fit.
There is no report for the time variation of flux density at
submillimeter wavelength in this galaxy so far.
In addition, the measurements for the total flux density in the plot
is distributed over three decades, but only $450~\mu$m shows the
large difference, which is unlikely to be caused by the time
variation.
We therefore assume that the JCMT SCUBA value is more accurate.
Using the fit, we estimated the total flux density at $435~\mu$m as
5.9~Jy.
This suggests that our $435~\mu$m continuum observations missed
$\sim58\%$ of the total flux density.

Since this missing flux is larger than our flux error of $\sim30\%$,
and the effect of the flux smearing due to phase or baseline errors
is small, this missing flux is probably due to the existence of an
extended component.
If the missing flux of 3.4~Jy is due to an extended component with a
Gaussian distribution that has a full width at half maximum of
$\sim3\arcsec$, its peak will be only twice of our r.m.s. noise.
We did not detect any significant signal with larger beam as
mentioned above, but the r.m.s.\ noise level increased in the larger
beam images due to a small number of data points at shorter
baselines.
It is therefore not clear whether the missing flux in our data is due
to the lack of shortest baselines or to low S/N.

\subsection{$^{12}$CO(6-5) Line Emission}
\label{res-co65}

\begin{figure}
\plotone{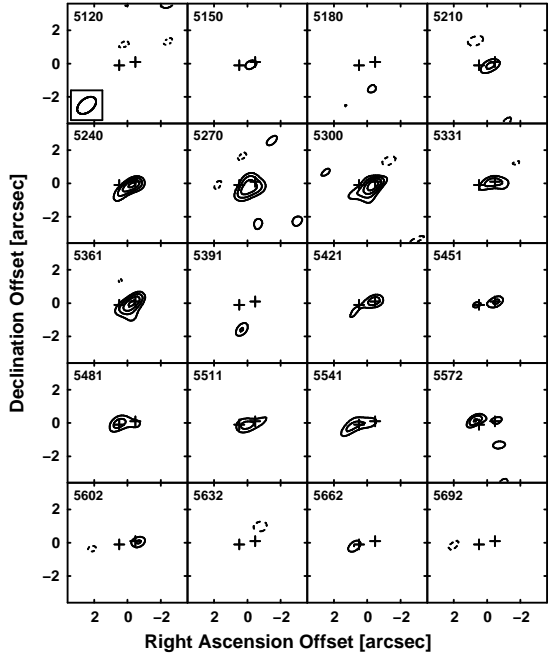}
\caption{
	Channel maps of the $^{12}$CO(6-5) line emission.
	Continuum emission is already subtracted.
	The LSR velocity (radio definition) in km s$^{-1}$ is shown at
	the upper-left corner of each channel map, and the synthesized
	beam ($1\farcs3\times0\farcs8$ or 500~pc $\times$ 310~pc) with
	the P.A.\ of $129\arcdeg$ is shown at the lower-left corner of
	the first channel map.
	The position offsets are measured from
	$\alpha(2000) = 15^{\rm h}34^{\rm m}57\fs25$ and
	$\delta(2000) = 23\arcdeg30\arcmin11\farcs4$.
	The two crosses in each channel map indicate the 1.3~mm continuum
	peaks \citep{sak99}.
	The contour levels are $-3, 3, 4, 5,$ and $6\sigma$,
	where $1\sigma$ = 535~mJy~beam$^{-1}$ (= 1.4 K).
\label{fig-co65ch}}
\end{figure}

\begin{figure}
\plotone{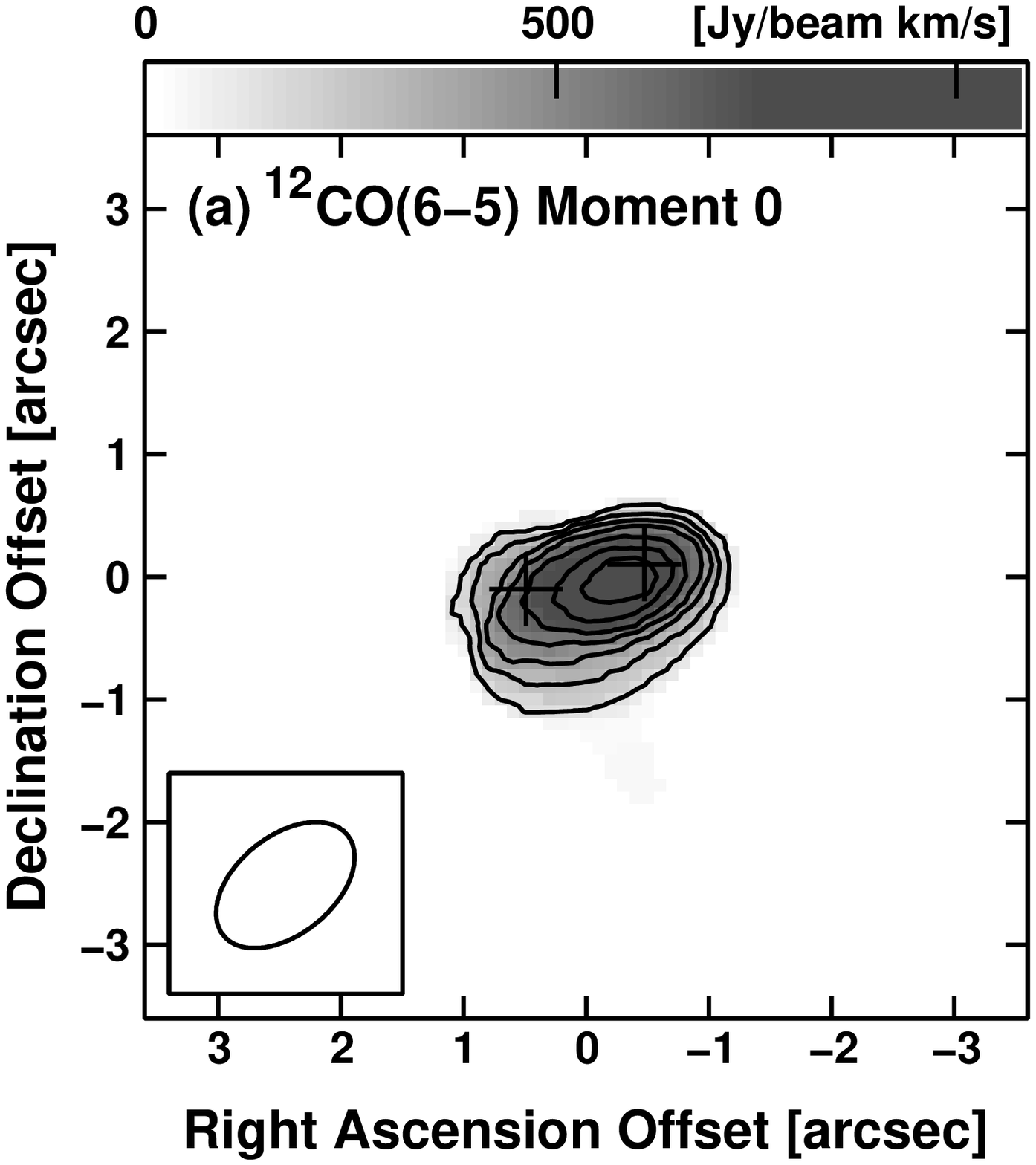}
\plotone{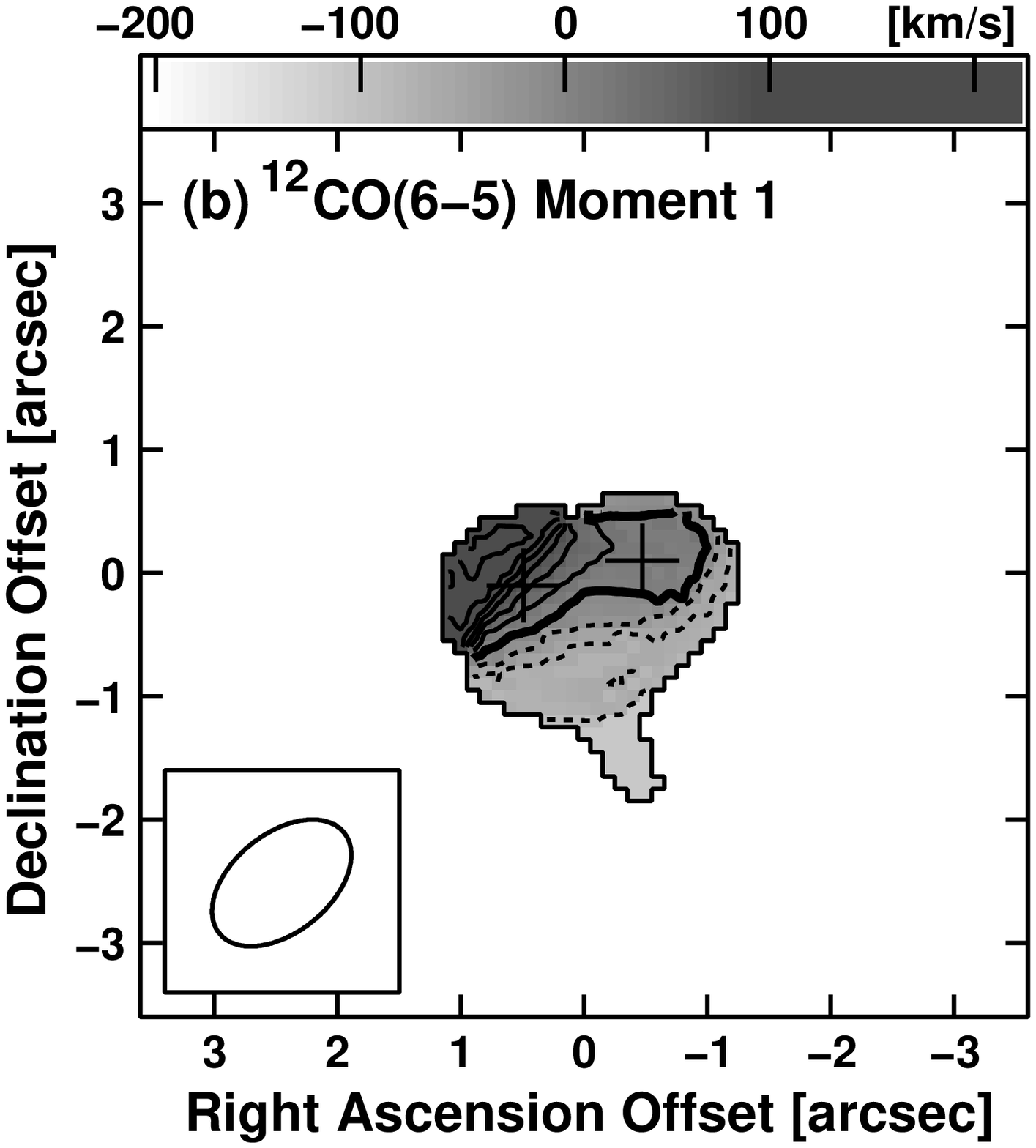}
\caption{
	(a) Integrated intensity (moment 0) and (b) intensity-weighted
	mean velocity (moment 1) maps of the $^{12}$CO(6-5) line
	emission.
	Continuum emission is already subtracted.
	The position offsets, the synthesized beam, and the crosses are
	the same as in Fig.~\ref{fig-co65ch}.
	The contour levels for the moment 0 map are (2, 4, 6, 8, 10, 12
	and $14)\times68$~Jy~beam$^{-1}$~km~s$^{-1}$ (= 175 K km s$^{-1}$).
	The contour levels for the moment 1 map are $-90, -60, -30, 0,
	30, 60, 90, 120, 150,$ and $180$~km~s$^{-1}$, where 0~km~s$^{-1}$
	corresponds to the LSR velocity of 5351~km~s$^{-1}$.
	Zero velocity is shown in thick solid contour, and negative and
	positive velocities are shown in dashed and solid contours,
	respectively.
\label{fig-co65mom01}}
\end{figure}

The channel maps of the $^{12}$CO(6-5) line emission at the nuclear
region of Arp 220 are shown in Fig.~\ref{fig-co65ch}, and the
integrated intensity (moment 0) and intensity-weighted mean velocity
(moment 1) maps are shown in Fig.~\ref{fig-co65mom01}.
The channel maps reveal that the molecular gas traced by the
$^{12}$CO(6-5) line exhibits different kinematics around each
nucleus; molecular gas associated with the eastern nucleus tends to
display gas kinematics moving from south-west to north-east, and that
associated with the western nucleus from south-east to north-west.
These features can also be seen in the moment 1 map for the eastern
nucleus.
Molecular gas around the western nucleus in the moment 1 map, on the
other hand, does not exhibit a clear velocity gradient as in the
channel maps.
This may be due to the nature of the moment 1 map (i.e., intensity
weighted velocity field map) with weak emission at the northwestern
region.
The kinematics on spatial scales larger than that immediately
associated with the individual nucleus shows a velocity gradient
along north-east to south-west direction with much shallower gradient
than that of the eastern nucleus.
All of these kinematics features mentioned above are also seen in the
previously published lower-J $^{12}$CO lines
\citep{sak99,sak08,dow98,sco97}.
The kinematics in the west nucleus displays a smaller velocity
gradient in our data, probably due to poorer spatial resolution and
lower S/N of our data.
The detection of the large-scale kinematics feature suggests that our
data detected some of the extended (a few arcsecond scale) component,
but much less than that detected in $^{12}$CO(1-0), $^{12}$CO(2-1),
or $^{12}$CO(3-2) observations \citep{sco97,dow98,sak99,sak08}.

The integrated intensity image, on the other hand, displays a single
peak that is elongated along the two nuclei, which looks different
from our $435~\mu$m continuum image or the previously published high
resolution $^{12}$CO(2-1) and $^{12}$CO(3-2) images
\citep{sak99,sak08,dow98} at first glance.
Our image is rather similar to the low resolution maps of
$^{12}$CO(2-1) line \citep{sak99,sco97} without diffuse and extended
(larger than a few arcsecond) emission from their maps.
The total integrated $^{12}$CO(6-5) intensity is
$1250\pm250$~Jy~km~s$^{-1}$ in our data.
Since there is no published result for the $^{12}$CO(6-5) line
emission, we could not estimate the missing flux of our
$^{12}$CO(6-5) data.

\begin{figure}
\epsscale{1.0}
\plotone{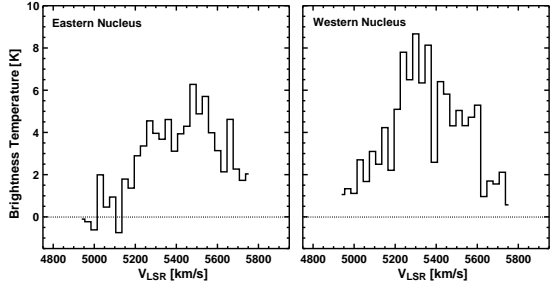}
\caption{
	Spectra of the $^{12}$CO(6-5) line at the two nuclei.
	Continuum emission is already subtracted.
\label{fig-co65spec}}
\end{figure}

Fig.~\ref{fig-co65spec} shows the $^{12}$CO(6-5) line spectrum at
each nucleus.
The peak brightness temperature of the western nucleus is
$8.7\pm2.6$~K ($3.4\pm1.0$~Jy~beam$^{-1}$) around the LSR velocity of
$\sim5300$~km~s$^{-1}$, and that of the eastern
nucleus $6.1\pm1.8$~K ($2.4\pm0.7$~Jy~beam$^{-1}$) around
$\sim5500$~km~s$^{-1}$.
The overall velocity ranges for the two nuclei are similar to those
in the high resolution $^{12}$CO(2-1) and $^{12}$CO(3-2) observations
\citep{sak99,sak08}.

\subsection{1.32~mm and 1.38~mm Continuum Emission}
\label{res-cont13}

\begin{figure}
\plotone{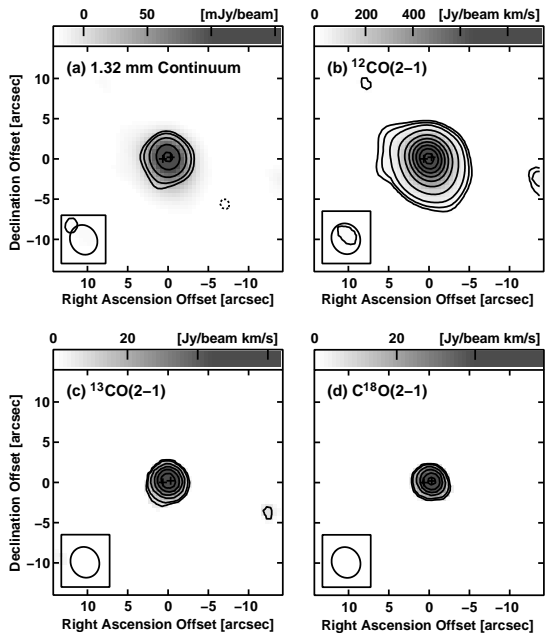}
\caption{
	Continuum and line images taken in the low frequency band.
	Continuum is already subtracted from the line images.
	Synthesized beams are shown at the lower-left corner of each
	image, and their sizes are in Table~\ref{tab-obs-contline}.
	The crosses are the same as in Fig.~\ref{fig-co65ch}.
	The position offsets are measured from
	$\alpha(2000) = 15^{\rm h}34^{\rm m}57\fs24$ and
	$\delta(2000) = 23\arcdeg30\arcmin11\farcs3$.
	(a) 1.32~mm continuum.
		Contour levels are $-3, 3, 5, 10, 20,$ and $30\sigma$, where
		$1\sigma$ = 4.8~mJy.
	(b) $^{12}$CO(2-1) integrated intensity.
		The contour levels are 5, 10, 20, 50, 100, 150, ...,
		$350\times2.5$~Jy~beam$^{-1}$~km~s$^{-1}$
		(= 4.8 K km s$^{-1}$).
	(c) $^{13}$CO(2-1) integrated intensity.
		The contour levels are 3, 5, 10, 15, 20, and
		$25\times2.1$~Jy~beam$^{-1}$~km~s$^{-1}$
		(= 4.0 K km s$^{-1}$).
	(d) C$^{18}$O(2-1) integrated intensity.
		The contour levels are the same as in the $^{13}$CO(2-1) map.
\label{fig-co21cont}}
\end{figure}

Both 1.32~mm (226.46~GHz) and 1.38~mm (216.46~GHz) continuum
emissions are detected at a significant signal level
($\sim30\sigma$), and the emission is unresolved at both wavelengths
at our resolution of $3''-4''$ (the 1.32~mm image is shown in
Fig.~\ref{fig-co21cont}a; the 1.38~mm image is almost the same as the
1.32~mm image, and not shown here).
The total flux densities for 1.32~mm and 1.38~mm are $167\pm33$~mJy
and $160\pm32$~mJy, respectively.
From the dust continuum fitting (Sect.~\ref{res-cont435}), we
estimate the total flux densities at 1.32~mm and 1.38~mm as 153~mJy
and 178~mJy (including the non-thermal flux contribution).
The observed total flux density therefore agrees with that from the
SED fitting within our calibration error.
We therefore conclude that our 1.32~mm and 1.38~mm continuum data
have no missing flux.

\subsection{$^{12}$CO, $^{13}$CO, and C$^{18}$O J=2-1 Line Emissions}
\label{res-co21}

The $^{12}$CO(2-1) line image exhibits an extended structure along
north-east and south-west direction even with our low resolution
image (Fig.~\ref{fig-co21cont}b), which is consistent with the past
interferometric maps \citep{sak99,dow98,sco97}.
On the other hand, the $^{13}$CO(2-1) and C$^{18}$O(2-1) line images
(Fig.~\ref{fig-co21cont}c, d) are unresolved at our resolution.

The integrated intensities of the $^{12}$CO, $^{13}$CO, and C$^{18}$O
J=2-1 lines are $1430\pm290$~Jy~km~s$^{-1}$,
$45.7\pm9.1$~Jy~km~s$^{-1}$, and $31.5\pm6.3$~Jy~km~s$^{-1}$,
respectively.
Our observation covers only $\sim72\%$ of the line width of the
C$^{18}$O(2-1) line (Sect.~\ref{obs}), so the derived value should be
a lower limit (see also Sect.~\ref{res-ratio}).

We compared the integrated intensities with the single dish results
for the $^{12}$CO(2-1) and $^{13}$CO(2-1) lines.
The JCMT observations of the $^{12}$CO(2-1) line
($21\arcsec-22\arcsec$ resolution) indicate
$1730\pm350$~Jy~km~s$^{-1}$ \citep{wie02} and
$1549\pm311$~Jy~km~s$^{-1}$ \citep{gre08}.
The flux differences between our and these observations are within
the calibration errors.
The $^{13}$CO(2-1) line was observed with the Institut de Radio
Astronomie Millimetrique (IRAM) 30~m telescope ($11\arcsec$
resolution) and JCMT ($21\arcsec$ resolution) \citep{gre08}, and they
obtained the integrated intensities of $60\pm13$~Jy~km~s$^{-1}$ and
$70\pm16$~Jy~km~s$^{-1}$, respectively.
The differences between our and these observations are again
explained by the calibration errors.
We therefore conclude that our $^{12}$CO(2-1) and $^{13}$CO(2-1)
line data have no missing flux.

Since there is no single dish C$^{18}$O(2-1) line observation, we
cannot estimate the missing flux for this line.
On the other hand, since we did not see any significant missing flux
in the 1.32~mm and 1.38~mm continuum data and the $^{12}$CO(2-1) and
$^{13}$CO(2-1) line data, we expect no significant missing flux in
the C$^{18}$O(2-1) line data.

\subsection{Line Spectra and Ratios}
\label{res-ratio}

\begin{figure}
\plotone{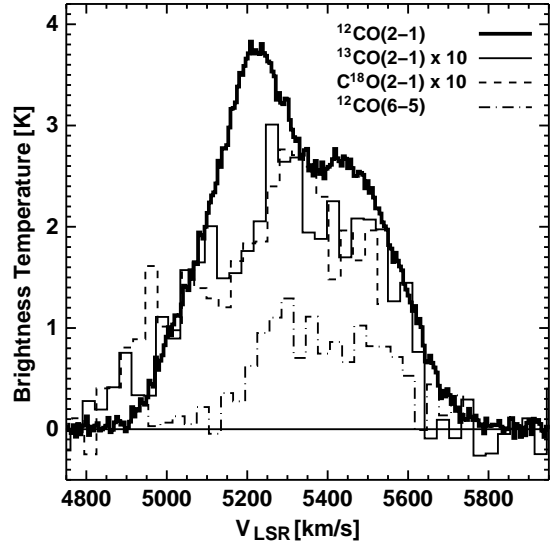}
\caption{
	Multiple CO line spectra toward the central region of Arp 220
	at $3\farcs9\times3\farcs5$ resolution (P.A.\ = $27\arcdeg$).
	Thick solid, thin solid, thin dashed, and thin dash-dot lines are
	the spectra of the $^{12}$CO(2-1), $^{13}$CO(2-1),
	C$^{18}$O(2-1), and $^{12}$CO(6-5) lines, respectively.
	Vertical axis is brightness temperature in Kelvin, and horizontal
	axis is LSR velocity in km~s$^{-1}$.
	Due to the weakness of the $^{13}$CO(2-1) and C$^{18}$O(2-1)
	lines, these intensities are increased by a factor of 10 in this
	figure.
	Since the C$^{18}$O(2-1) line is located at the edge of the
	bandpass, the spectrum finishes around the LSR velocity of
	5550~km~s$^{-1}$.
	Note that the $uv$ coverage is different between J = 2 -- 1 lines
	and the J = 6 -- 5 line in this figure.
\label{fig-cospec}}
\end{figure}

To compare the line spectra and intensities of all the four obtained
lines, we convolved the data into the largest synthesized beam size
of $3\farcs9\times3\farcs5$ with the P.A.\ of $27\arcdeg$, which is
the beam size of the $^{13}$CO(2-1) and C$^{18}$O(2-1) lines.
Fig.~\ref{fig-cospec} shows the spectra of all these four lines.
Note that the sensitivity toward extended structure (i.e., $uv$
coverage) is different between the J = 2 -- 1 lines and the
J = 6 -- 5 line in this figure.

The $^{12}$CO(2-1) line shows a double peak profile with stronger
intensity at lower velocity, which is consistent with the past
single dish measurements \citep{wie02,sol90}.
The line width of the $^{12}$CO(2-1) line reaches 900~km~s$^{-1}$
(4900 -- 5800~km~s$^{-1}$).
The $^{13}$CO(2-1) and C$^{18}$O(2-1) lines exhibit very similar line
profiles and intensities.
The $^{12}$CO(6-5) line is mostly emitted at higher velocities and is
weak at lower velocities.
This asymmetry is similar to that of the HCN(4-3) line \citep{wie02}.

We matched the shortest $uv$ distance between the $^{12}$CO(6-5) and
$^{12}$CO(2-1) data to match the spatial structures between these two
lines, and measured the $^{12}$CO(6-5)/(2-1) intensity ratio to be
$0.34\pm0.12$.
This value is much lower than the lower-J ratios, such as
$^{12}$CO(3-2)/(2-1) of $0.85\pm0.24$ \citep{wie02} or
$^{12}$CO(2-1)/(1-0) of $0.65\pm0.1$ \citep{sco97}.
This is mostly due to the different line profile of the
$^{12}$CO(6-5) line from the lower-J CO lines.
This indicates that the lower velocity is dominated by lower-J CO
lines, but the higher velocity is rich in higher-J lines.
This trend is consistent with the observations of \citet{wie02} that
the $^{12}$CO(3-2) line intensity decreased relative to the
$^{12}$CO(2-1) line in the lower velocity part, but stay almost
constant at higher velocities.
The relation between the spatial distribution and the velocities of
the molecular gas in this galaxy is, however, not simple;
all the molecular gas components in this galaxy, namely the two
nuclei and the extended component, have low and high velocities
(Fig.~\ref{fig-co65spec}; see also \citealt{sak99,sak08}).
It is therefore difficult to tell which component contributes to high
or low excitation condition only from the large-scale line spectra.
The $^{12}$CO transition ratios for each nucleus is derived in
Sect.~\ref{dis-co}.

The $^{12}$CO(2-1)/$^{13}$CO(2-1) line ratio is $13.0\pm3.7$ at our
resolution of $3\farcs9\times3\farcs5$.
We also convolved our data to $13\arcsec$ resolution ($\approx$
single dish resolution), and the ratio was $16.2\pm4.6$.
These values are similar or slightly lower than the values observed
in U/LIRGs, and similar or somewhat higher than starburst or Seyfert
galaxies;
the $^{12}$CO(2-1)/$^{13}$CO(2-1) line ratios in U/LIRGs observed
with single dish telescopes are $23.5\pm4$ \citep{cas92} and $16\pm8$
\citep{gle01}, and those in starburst and Seyfert galaxies are
$13\pm5$ \citep{aal95} and $13\pm1$ \citep{pap98}, respectively.

The line profiles and the line intensities of the $^{13}$CO(2-1) and
C$^{18}$O(2-1) are almost identical (Fig.~\ref{fig-cospec}), and the
$^{13}$CO(2-1)/C$^{18}$O(2-1) line intensity ratio is $1.0\pm0.3$
between the velocity range of 4800~km~s$^{-1}$ and 5550~km~s$^{-1}$,
which is the velocity range we observed the C$^{18}$O(2-1) line
(Sect.~\ref{obs}).
The $^{13}$CO(2-1) line has 92\% of its total integrated intensity in
this velocity range, so the result with the full line width will not
change significantly.
This very low $^{13}$CO(2-1)/C$^{18}$O(2-1) intensity ratio matches
very well with $^{13}$CO(1-0)/C$^{18}$O(1-0) $= 1.0\pm0.3$
\citep{gre08}.
We discuss this ratio in Sect.~\ref{dis-abn}.

\section{DISCUSSION}
\label{dis}

\subsection{Dust Emissivity Index and Dust Opacity
	of the Two Nuclei}
\label{dis-sed}

\begin{deluxetable*}{cccc}
\tablecaption{Fitting results of dust spectral energy distributions
	\label{tab-dust}}
\tablehead{
	\colhead{Source} & \colhead{Dust Temperature $T_{\rm d}$}
	& \colhead{Emissivity Index $\beta$}
	& \colhead{Critical Frequency (Wavelength) $\nu_{\rm c}$} \\
	& \colhead{[K]} & & \colhead{[GHz ($\micron$)]}
	}
\startdata
Arp 220 (total)    & $51- 66$ & $1.3-1.4$ & $2000-2200$ ($\sim140- 150$) \\
East Nucleus       & $49-120$ & $1.8-2.1$ &  $370- 520$ ($\sim580- 810$) \\
West Nucleus       & $97-310$ & $0.7-1.2$ &  $190- 610$ ($\sim490-1600$) \\
Extended Component & $\sim38$ & $\sim2.4$ &  $\sim1500$ ($\sim200$)
\enddata
\end{deluxetable*}

\begin{figure}
\plotone{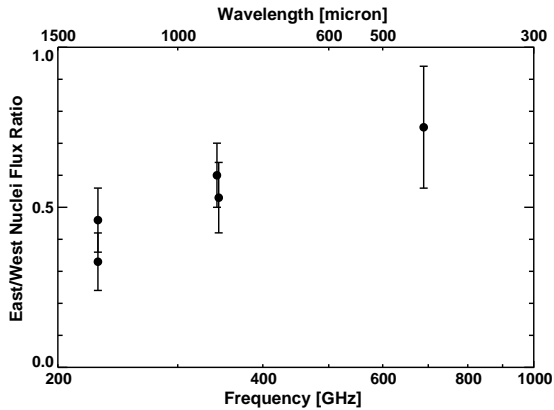}
\caption{
	Continuum flux ratio between the eastern and western nuclei as a
	function of frequency.
	Our data point is the right-most one, and the data for 229~GHz
	and 343~GHz are the same as in Fig.~\ref{fig-sed}.
\label{fig-fratio}}
\end{figure}

We plot in Fig.~\ref{fig-fratio} the flux ratios between the two
nuclei as a function of frequency.
Although our continuum data have $\sim30\%$ of calibration errors and
a large amount ($\sim60\%$) of missing flux,
the flux density ratio between the two nuclei is accurate, since it
depends on the noise level of the map.
Hence this flux ratio diagram has higher accuracy than the SED
diagrams shown in Fig.~\ref{fig-sed}.
The flux ratio in our data is $0.75\pm0.19$.
If we assume that the missing flux of our data is due to the extended
emission (see Sect.~\ref{res-cont435}), the size of the extended
emission is larger than the separation of the two nuclei.
Therefore the correction of the missing flux will increase both the
fluxes of the two nuclei with almost the same amount, and the flux
ratio increases toward unity, namely the ratio derived above is the
lower limit.
Contamination of the flux from one nucleus to the other due to the
large beam size is small; this effect lowers the ratio by $\sim6\%$,
which is smaller than the calibration errors.
As frequency goes down to $\sim200$~GHz, the flux ratio also goes
down, and the flux of the eastern nucleus is about half or less of
that of the western nucleus.
Since the data for the two lower frequencies may not have significant
missing flux \citep[e.g.,][]{sak99,sak08}, the difference between the
ratio at 689~GHz and those at lower frequencies will be more
pronounced if we correct for the missing flux of the 689~GHz data.
Since the emission at these frequencies is dominated by dust
emission, this result suggests that the dust SEDs are different
between the two nuclei.

We then made continuum SEDs for the two nuclei at submillimeter
wavelength (Fig.~\ref{fig-sed}) as well as that for the total flux
density to discuss the dust property differences more quantitatively.
The crosses, filled squares, and filled diamonds in the plot are the
data for the total flux density, east nucleus, and west nucleus,
respectively.
The solid, dashed, and dotted lines are the $\chi^2$ fitting of the
data with a function $\epsilon B(T_{\rm d})$, where $B(T_{\rm d})$ is
the Planck function for temperature $T_{\rm d}$ and $\epsilon$ is the
emissivity function.
We adopted a form $\epsilon = 1-\exp[-(\nu/\nu_{\rm c})^\beta]$;
$\nu_{\rm c}$ is the critical frequency where the opacity is unity,
and $\beta$ is the emissivity index.
We adopted the source sizes of $0\farcs27 \times 0\farcs14$ for the
eastern nucleus and $0\farcs16 \times 0\farcs13$ for the western
nucleus \citep{sak08}, since their images have the highest spatial
resolution at the highest frequency around submillimeter wavelengths.
For the source size of the total flux density, we adopted the
deconvolved size of the extended $^{12}$CO(2-1) emission of
$1\farcs94 \times 1\farcs28$ \citep{sco97}.
The flux density for the non-thermal component is subtracted from all
the fluxes in the plots before the fitting, following the method
explained in \citet{sco91}, although it is not significant especially
at high frequencies; about several milli-Jy or several percent of the
flux around $200-300$~GHz, and even smaller at $600-700$~GHz.
We also estimated the CO flux contamination into the total flux, but
most of the data are affected for only a few \%, and the fitting
result did not change.
The fitting results are summarized in Table~\ref{tab-dust}.

We obtained from the fit to the total flux density $T_{\rm d}$ of
51~K and $\beta$ of 1.4, which are consistent with the past estimates
of $T_{\rm d} \sim 40-60$~K with $\beta \sim 1.2-2.0$
\citep{sco91,dow98,dun01,kla01}.
The critical frequency is estimated to be 2200~GHz
($\sim140~\micron$), which is also consistent with the estimation
that dust is already optically thick at $\sim100~\micron$
\citep{sco91,kla01}.
If we adopt a smaller source size for the total flux of
$1\farcs13$ \citep{sco97}, $T_{\rm d}$, $\beta$, and $\nu_{\rm c}$
would be 66~K, 1.3, and 2000~GHz ($\sim150~\micron$), respectively.
The temperature is slightly higher, but the critical frequency and
the emissivity index are still consistent with the past estimates.

The fitting for the two nuclei gives high temperatures of 83~K and
180~K for the east and west nuclei, respectively.
These high temperatures are due to the small source sizes.
The emissivity indices are different between the two nuclei, 2.1 for
the eastern nucleus and 1.2 for the western nucleus.
The critical frequencies are estimated to be 370~GHz
($\sim810~\micron$) for the eastern nucleus and 190~GHz
($\sim1.6$~mm) for the western nucleus.

The large uncertainties in our fitting for the two nuclei are the
source flux density and the adopted source size.
If we increase the 689~GHz flux density of each nucleus by 30\%,
which corresponds to the calibration error of our data may have
(Sect.~\ref{obs}), the fitting results would be $T_{\rm d}=120$~K
and 310~K, $\beta=1.8$ and 0.7, and $\nu_{\rm c}=520$~GHz
($\sim580~\micron$) and 610~GHz ($\sim490~\micron$) for the eastern
and western nuclei, respectively.
We adopted the source size derived at $860~\micron$ \citep{sak08},
but the effective source size at $435~\micron$ may be different from
that at $860~\micron$.
This is because the opacity is wavelength dependent,
and therefore the effective source size is also wavelength dependent.
Ideally we need to determine the source size at each wavelength, but
our spatial resolution at $435~\micron$ is too low for this.
Since opacity is higher at shorter wavelength, the effective source
size at $435~\micron$ may be larger than what we adopted.
If we increase the source size (area) of each nucleus by 50\%, which
roughly corresponds to the deconvolved size of the double nucleus in
our data, the fitting results would be $T_{\rm d}=49$~K and 97~K,
$\beta=2.1$ and 1.1, and $\nu_{\rm c}=400$~GHz ($\sim750\micron$) and
210~GHz ($\sim1.4$~mm) for the eastern and western nuclei,
respectively.

The fitting results, with the uncertainties, indicate that the dust
temperature, the emissivity index, and the critical frequency for the
eastern nucleus are better constrained than the past observations,
because of our high frequency observations with high spatial
resolution.
The eastern nucleus seems to have a warm temperature of 49 -- 120~K,
a steep emissivity index of $\sim2$, and
becomes optically thick at frequencies above $\sim400$~GHz.
\citet{sak08} derived a dust temperature of 30 -- 160~K, so our
estimate narrows the range.
They also estimated the 350~GHz opacity of 2.8 for $\beta=2$ (and 0.8
for $\beta=1$), hence our result is somewhat lower, but still both
results indicate a high opacity condition at submillimeter
wavelengths in the eastern nucleus.

The emissivity index and the critical frequency for the western
nucleus are less constrained than those for the eastern nucleus, but
are better constrained than the past observations.
Our fitting results indicate a shallow emissivity index of about
unity and a low critical frequency of $\lesssim600$~GHz.
\citet{dow07} estimated the 230~GHz opacity as $\ge0.7$ and
\citet{sak08} estimated the 350~GHz opacity as $0.8-5.3$ for
$\beta=2$ (the estimated ranges depends on the source size and the
flux errors).
Both of these observations and our results indicate the western
nucleus is optically thick at submillimeter wavelengths.

The fitted temperature for the western nucleus, on the other hand,
has a large range, and therefore it does not set a tighter limit than
the past estimations.
It appears that the western nucleus is warmer than the eastern
nucleus, with dust temperatures of a few hundred Kelvin.
\citet{dow07} suggested a dust temperature of 170~K, and
\citet{sak08} derived a temperature of 90 -- 180~K, hence our
estimate is consistent.

Our results indicate that the derived properties, especially the
emissivity, indices are different between the two nuclei, suggesting
that the dust properties, such as dust size distributions or dust
compositions, are different.
This difference may reflect the dust properties of the original host
galaxies of each nucleus, or the difference in activities, such as
star formation or AGN activities, in the nuclei after the merger.

\subsection{Extended Component in the Dust Emission}
\label{dis-ext}

Our $435~\mu$m (689~GHz) continuum data missed a significant amount
($\sim61\%$) of the total flux.
Here we discuss whether this missing flux can be attributed to an
extended component in the dust emission.
The molecular gas clearly has an extended component with a size of
$\sim2\arcsec$ ($\sim1$~kpc), which is interpreted as a molecular gas
disk from the gas kinematics \citep{sco97,dow98,sak99}.
In dust emission, on the other hand, the extended component is weakly
detected or not detected at a significant signal level.
At 1.3~mm, \citet{dow98} suggested that the flux of $55\pm11$~mJy can
be attributed to the flux from the extended component, but
\citet{sak99} did not detect any significant emission from the
extended component.
At $860~\micron$, \citet{sak08} attributed about 25\% of the
total flux density emission from outside the two nuclei.

\begin{deluxetable*}{ccccccc}
\tabletypesize{\scriptsize}
\tablecaption{Integrated $^{12}$CO intensities and intensity ratios
	at each nucleus
	\label{tab-co}}
\tablehead{
	\colhead{Source} %& \colhead{Peak $435~\mu$m Flux}
	& \colhead{I($^{12}$CO 2-1)}
	& \colhead{I($^{12}$CO 3-2)}
	& \colhead{I($^{12}$CO 6-5)}
	& \colhead{$^{12}$CO(6-5)/(2-1)}
	& \colhead{$^{12}$CO(6-5)/(3-2)}
	& \colhead{$^{12}$CO(3-2)/(2-1)} \\
	& \colhead{[K km s$^{-1}$]}
	& \colhead{[K km s$^{-1}$]}  & \colhead{[K km s$^{-1}$]} & & &
	}
\startdata
East Nucleus %& $1.28 \pm 0.26$
             & $5540 \pm 830$  & $5040 \pm 760$  & $2390 \pm 720$
             & $0.43 \pm 0.14$ & $0.47 \pm 0.16$ & $0.91 \pm 0.19$ \\
West Nucleus %& $0.96 \pm 0.19$
             & $4460 \pm  670$  & $5140 \pm 770$  & $2840 \pm 850$
             & $0.64 \pm 0.21$ & $0.55 \pm 0.19$ & $1.15 \pm 0.24$
\enddata
\tablecomments{The adopted spatial resolution is
	$1\farcs3\times0\farcs8$ with P.A.\ of $129\arcdeg$.
	The $^{12}$CO(6-5) data for the eastern and western nuclei of
	Arp 220 are our data, and those for $uv$-matched
	$^{12}$CO(2-1) and $^{12}$CO(3-2) data are from \citet{sak99} and
	\citet{sak08}.}
\end{deluxetable*}

\begin{figure*}
\plottwo{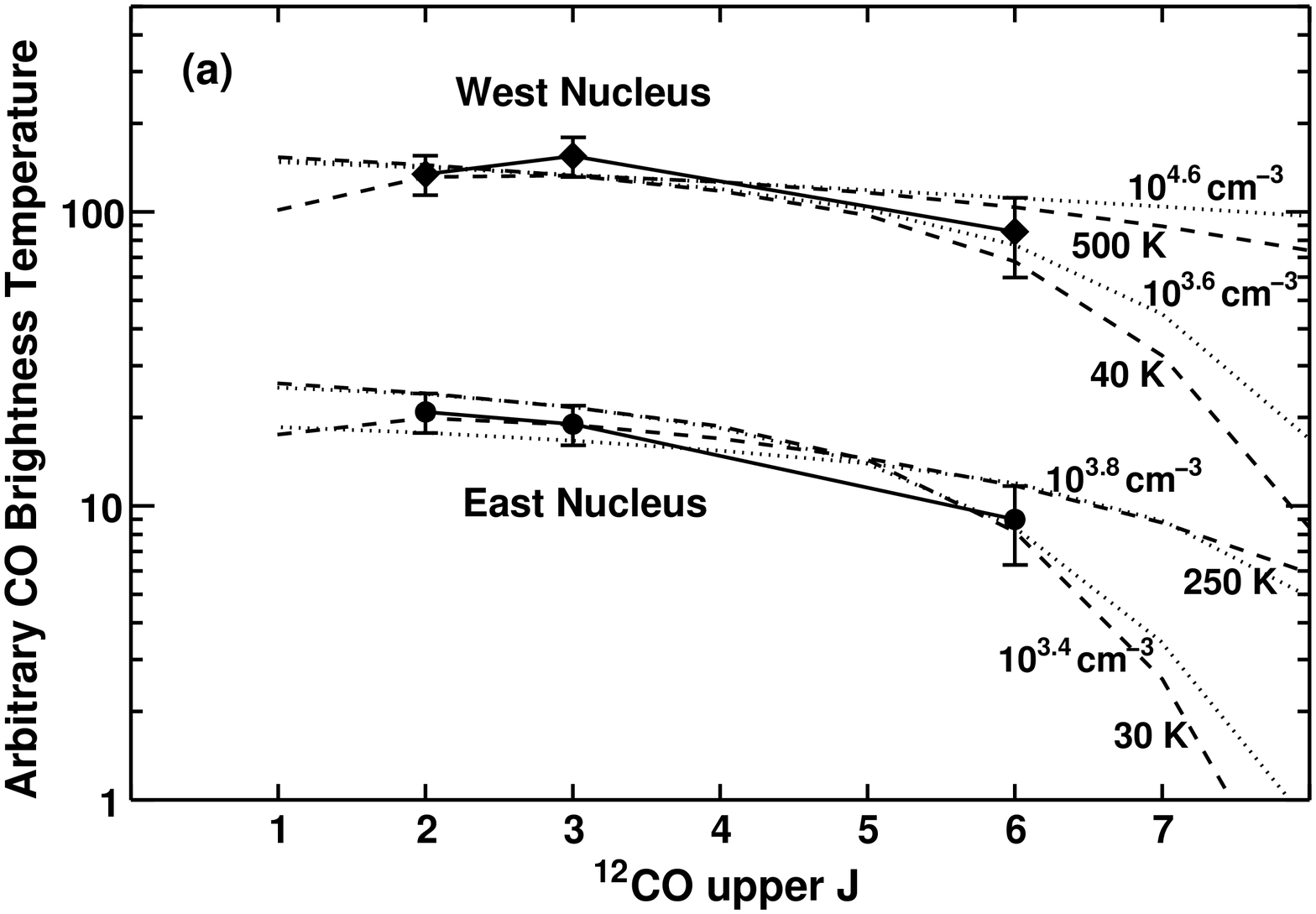}
	{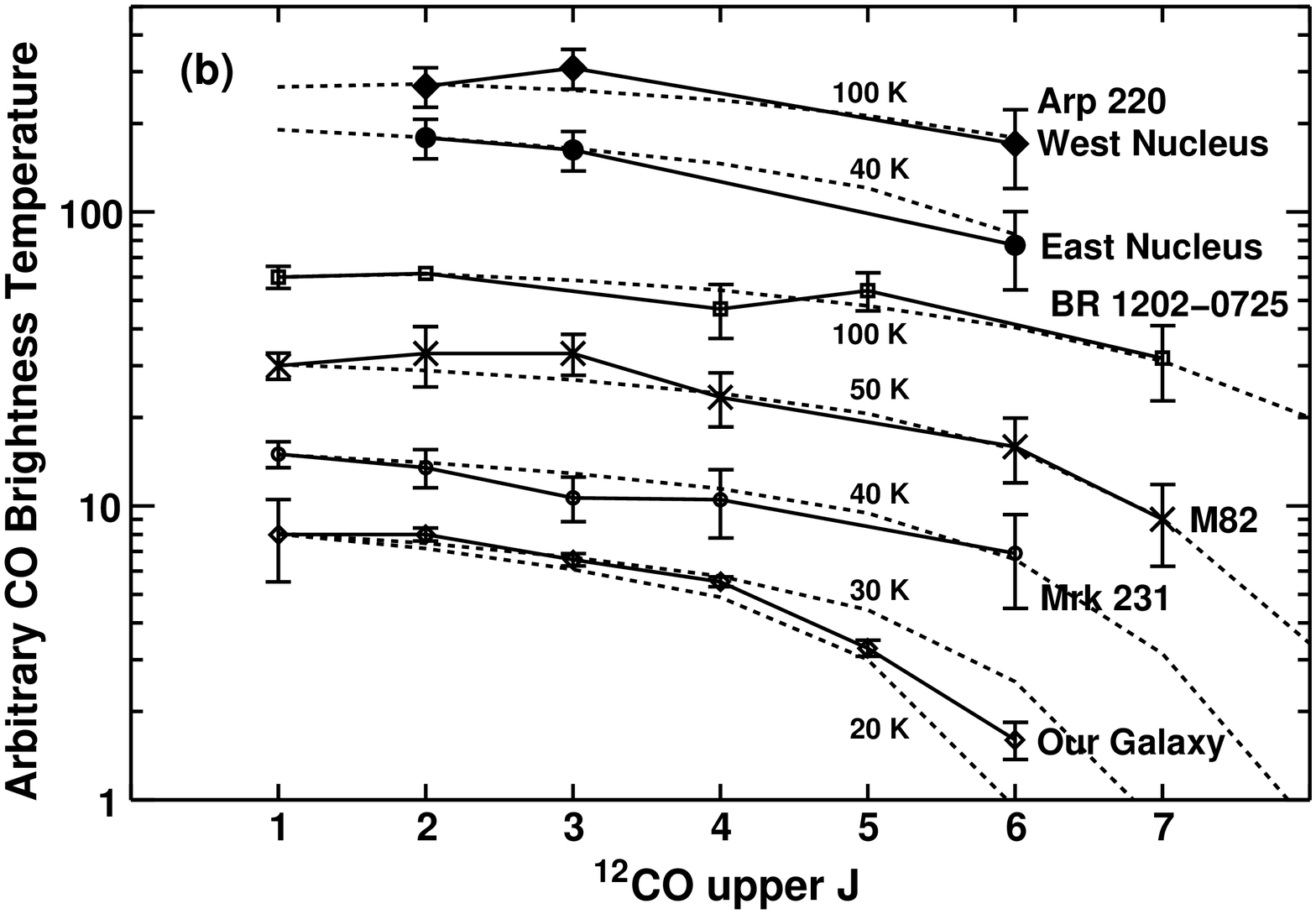}
\caption{Rotational transition dependence of the $^{12}$CO brightness
	temperatures (SEDs) of the double nucleus in Arp 220 and those of
	other galaxies.
	The horizontal axis is the upper rotational transition levels of
	the $^{12}$CO lines and the vertical axis is the brightness
	temperatures of the $^{12}$CO lines on an arbitrary scale.
	We fixed the $Z$($^{12}$CO)/($dv/dr$) of
	$5\times10^{-5}$~(km~s$^{-1}$~pc$^{-1}$)$^{-1}$.
	(a) Arp 220 $^{12}$CO SED of each nucleus overplotted with the
		LVG calculation results.
		Dashed lines are for two temperatures at a density of
		$10^{3.6}$~cm$^{-3}$.
		Dotted lines are for two densities at a temperature of 50~K.
	(b) $^{12}$CO SED of Arp 220 and other galaxies.
		The multi-J $^{12}$CO data of the Galactic Center
		($|l|<2.5\arcdeg$) are taken from \citet{fix99}, and those of
		M82 and Mrk 231 are compiled by \citet{wei05a} and
		\citet{pap07}, respectively.
		The $^{12}$CO data of BR 1202--0725 are taken from the
		following papers:
		  J=1-0: \citet{rie06},
		  J=2-1: \citet{car02},
		  J=4-3, J=7-6: \citet{omo96}, and
		  J=5-4: \citet{oht96}.
		The best fitted molecular gas temperature curves, made from
		LVG calculations for a density of $10^{3.6}$~cm$^{-3}$, are
		also overplotted for reference.
\label{fig-coj}}
\end{figure*}

Assuming that all of the missing flux of our $435~\mu$m continuum
data is from the extended component, we derived the dust properties
of this extended component using the $\chi^2$ fitting mentioned above
using the data between 1.3~mm and $435~\mu$m.
We adopt for the size of this component the same size that we adopted
for the fitting of the total flux density, which is
$1\farcs94\times1\farcs28$ \citep{sco97}.
The fitting result is shown in Fig.~\ref{fig-sed} with a dash-dotted
line, and the derived values are $T_{\rm d}\sim38$~K, $\beta\sim2.4$,
and the $\nu_{\rm c}\sim1500$~GHz ($\sim200~\mu$m).
These values are roughly consistent with the values derived
by \citet{gon04} using ISO LWS data; they derived for the extended
component $T_{\rm d}\sim50$~K, $\nu_{\rm c}\sim3000$~GHz
($\sim100~\mu$m), and a source size of $\sim1\farcs8 - 1\farcs9$
(these values vary depending on their models) assuming $\beta$ of
2.0.
These results suggest that a significant amount ($>50\%$) of
$435~\mu$m flux is in the extended component as mentioned in
Sect.~\ref{res-cont435}, contrary to the longer wavelengths where
continuum is dominantly from the two nuclei.

A recent SMA U/LIRG survey shows that many of the sample galaxies
(9 out of 14) have large missing fluxes (typically $50-80\%$) in
$880~\mu$m continuum emission, even though many of the galaxies also
show compact distributions.
This result suggests that many of ULIRGs have extended continuum
emission with very luminous compact cores \citep{wil08}, similar to
the results of Arp 220.

\subsection{Excitation Conditions of Molecular Gas in the Two Nuclei}
\label{dis-co}

We made $^{12}$CO SEDs of the two nuclei using our $^{12}$CO(6-5)
data with the interferometric $^{12}$CO(2-1) and $^{12}$CO(3-2) data
\citep{sak99,sak08} to study the $^{12}$CO excitation conditions.
We first matched the shortest $uv$ distance for all three data sets,
and imaged at the same synthesized beam size of
$1\farcs3 \times0\farcs8$ with P.A.\ of $129\arcdeg$, which is the
same spatial resolution as our $^{12}$CO(6-5) image.
The integrated intensities and line ratios of these three lines for
each nucleus are listed in Table~\ref{tab-co}, and the relative
$^{12}$CO intensities of various transitions are shown in
Fig.~\ref{fig-coj} for each nucleus.

The decrease of $^{12}$CO intensities toward higher-J in the western
nucleus is smaller than that in the eastern nucleus, and indeed the
intensity ratios are higher for the western nucleus than those for
the eastern nucleus.
These differences are, however, within observational errors.
Thus the difference in the excitation conditions of molecular gas
between the two nuclei is not significant.

To discuss the excitation conditions more quantitatively, we
estimated the excitation conditions of the molecular gas using the
large-velocity-gradient (LVG) approximation \citep{gol74}.
The collision rates for CO of $\leq100$~K were taken from
\citet{flo85} and of $\geq250$~K were from \citet{mck82}.
In these calculations, we assume that all the $^{12}$CO emission
comes from the same region (i.e., one-zone model), and assume the
$^{12}$CO relative abundance over velocity gradient,
$Z$($^{12}$CO)/($dv/dr$), of
$5\times10^{-5}$~(km~s$^{-1}$~pc$^{-1}$)$^{-1}$.
In Fig.~\ref{fig-coj}(a), we overplotted two kinds of curves, one
(dashed lines) is to see the temperature dependence (we fixed
$n_{\rm H_{2}}$ of $10^{3.6}$~cm$^{-3}$), and another (dotted lines)
is to see the number density dependence (we fixed $T_{\rm k}$ of
50~K).
We only plotted the curves close to upper or lower limits to show
the possible ranges of the excitation conditions and the goodness of
the fitting.
Under the above LVG assumptions, it is estimated that the eastern
nucleus has a molecular gas temperature of $\sim30-250$~K, or a
density of $\sim10^{3.5\pm0.2}$~cm$^{-3}$.
For the western nucleus, we could only derive the lower limits,
which are $\gtrsim40$~K for temperature and
$\gtrsim10^{3.5}$~cm$^{-3}$ for density.
As mentioned above, the estimated molecular gas conditions for the
two nuclei overlap, but the western nucleus tends to have higher
temperature or density than the eastern nucleus.
This tendency is similar to that derived from the dust SED fitting
(Sect.~\ref{dis-sed}).
Indeed, the derived molecular gas and dust temperatures for both
nuclei match well.
This suggests that both the molecular gas and dust reside at similar
regions and in thermal equilibrium.

We also calculated the dependence of our results on
$Z$($^{12}$CO)/($dv/dr$).
If we decrease $Z$($^{12}$CO)/($dv/dr$) by an order of magnitude
(i.e., decrease the $^{12}$CO relative abundance or increase the
velocity gradient or both; we fixed $n_{\rm H_{2}}$ of
$10^{4.0}$~cm$^{-3}$ or $T_{\rm k}$ of 100~K), the temperature or the
density increases by about a factor of a few in both nuclei.
If we increase $Z$($^{12}$CO)/($dv/dr$) by an order of magnitude (we
fixed $n_{\rm H_{2}}$ of $10^{3.4}$~cm$^{-3}$ or $T_{\rm k}$ of
30~K),
the temperature or the density decreases by about a factor of a few.
Therefore a small (within an order of magnitude) change in
$Z$($^{12}$CO)/($dv/dr$) does not change the result significantly.

Our $^{12}$CO SEDs are compared with those of other galaxies in
Fig.~\ref{fig-coj}(b).
We plotted the multi-J $^{12}$CO intensities of the Galactic Center
(normal and quiescent galaxy), M82 (nearby starburst galaxy), Mrk 231
(evolved ULIRG with an AGN at the nucleus), and BR 1202--0725
(radio-quiet and CO bright high-z quasar at z of 4.69)
together with the $^{12}$CO SEDs for the Arp 220 nuclei.
The brightness temperatures of the Galactic Center quickly decrease
with the increase of rotational transitions, but those of
BR 1202--0725, M82, and Mrk 231 stay almost constant even at high-J
transitions.
We overplotted the best fit temperature curves on each source for
comparison (as shown in Fig.~\ref{fig-coj}a, higher temperature can
be replaced with higher density).
The Galactic Center can be modeled well with low temperature (or low
density) conditions, but other galaxies are explained with higher
temperatures (or higher densities).
The two nuclei of Arp 220 is similar to these higher temperature
(density) galaxies, and different from the Galactic Center.
This suggests that the molecular gas excitation condition in the
double nucleus of Arp 220 is similar to these galaxies.
Note that the $^{12}$CO SED up to J=6-5 or 7-6 data with the current
accuracy are not enough to distinguish the excitation conditions
between these high temperature (density) galaxies, including the two
nuclei of Arp 220.
Higher accuracy or higher-J observations are needed to differentiate
the excitation conditions of these galaxies.

It is known that Arp 220, M82, and BR 1202--0725 have active star
formation inside.
The excitation conditions in the two nuclei of Arp 220 and in M82 are
averaged over the central a few hundred pc, and that in BR 1202--0725
is averaged over several kpc.
M82 has a gradient in the physical conditions from the center to the
outer region \citep{pet00}, and the physical conditions derived above
is more similar to those of the center, where the starburst region
exist.
BR 1202--0725 has two sources, north and south, and the southern
source may consist of two sources \citep{car02}, probably interacting
with each other.
The similar excitation conditions of molecular gas regardless of the
observed regions suggests that the observed molecular gas is biased
toward the gas closely related to the star forming regions, and the
effect of star forming activities to the exciting conditions of
surrounding molecular gas is similar.

The $^{12}$CO SED study can also be a useful tool to search for
AGN(s), since the nearby Seyfert galaxies exhibit strong enhancement
of higher-J $^{12}$CO lines toward AGNs \citep[e.g.,][]{mat04,hsi08}.
Fig.~\ref{fig-coj}(b) exhibits, however, that the $^{12}$CO SED of
the AGN hosting ULIRG Mrk 231 does not display any higher-J
enhancement, and $^{12}$CO SED comparison between Mrk 231 and the
two nuclei of Arp 220 shows no clear difference.
In addition, the comparison between Mrk 231, starburst dominated
galaxies M82 and BR 1202--0725 also show no clear difference.
We therefore could not find any evidence of an AGN in Arp 220 from
this $^{12}$CO SED study.
These results suggest that the AGN contribution to the surrounding
molecular gas (at least for Mrk 231) is much smaller than the
nearby Seyferts, possibly due to the smoothing effect by a larger
(linear scale) beam or to a larger opacity effect toward the AGN.

\subsection{Molecular Gas Abundance Anomaly in the Central Region of
	Arp 220?}
\label{dis-abn}

As mentioned in Sect.~\ref{res-ratio}, we obtained a very low
$^{13}$CO(2-1)/C$^{18}$O(2-1) line intensity ratio of about unity.
Recent SMA observations toward nearby active star forming galaxies
(NGC 253, NGC 1365, and NGC 3256) show $^{13}$CO(2-1)/C$^{18}$O(2-1)
ratios of $\sim4$ \citep{sak06a,sak06b,sak07}.
The $^{13}$CO(1-0)/C$^{18}$O(1-0) line ratios in nearby galaxies are
$\sim4$ \citep{sag91}, similar to the J=2-1 ratios.
If both the $^{13}$CO and the C$^{18}$O lines are optically thin,
the $^{13}$CO/C$^{18}$O line ratios for J=2-1 and J=1-0 are expected
to have almost the same values.
Some of interferometric $^{13}$CO and C$^{18}$O observations of
nearby galaxies show $^{13}$CO/C$^{18}$O ratio of about 2
\citep{mei04,cho07}, but not unity as in Arp 220 (note that some
regions observed by \citeauthor{mei04} show the $^{13}$CO/C$^{18}$O
ratios of about unity, but the S/Ns are low).
The intensity ratio of about unity in Arp 220 is therefore unusual
compared with those in other galaxies.

The intensity ratio may be closely related to the abundance ratio;
the intensity ratio is expected to be similar to the abundance ratio,
if both lines are optically thin.
The abundance ratio between $^{13}$CO and C$^{18}$O in our Galaxy is
5.5 for the Solar System and 12.5 for the Galactic Center, and that
in external galaxies is $3-5$ \citep{hen93}, assuming
  [$^{13}$CO]/[C$^{18}$O]
    = [$^{13}$C]/[$^{12}$C] $\times$ [$^{16}$O]/[$^{18}$O].
Indeed, the abundance ratios and the aforementioned intensity ratios
for external galaxies are similar.
The intensity ratio of about unity is unusual also from the abundance
viewpoint.

Here we discuss possible reasons for this low intensity ratio using
the LVG calculations.
We assume both $^{13}$CO and C$^{18}$O molecules are located at the
same region (one-zone model).
Note that since the brightness temperatures are different between
these two lines and the $^{12}$CO(2-1) line (see
Fig.~\ref{fig-cospec}), it is evident that these two lines and
the $^{12}$CO(2-1) line emanate from different regions.
We also assume that the [$^{13}$CO]/[C$^{18}$O] relative abundance
ratio of 4 \citep{wan04}.
Under this relative abundance ratio, both lines have to be optically
thick for the line ratio to be unity.
We calculated assuming $Z$($^{13}$CO)/($dv/dr$) of $1\times10^{-5}$,
$1\times10^{-6}$, and
$1\times10^{-7}$~(km~s$^{-1}$~pc$^{-1}$)$^{-1}$.
$Z$($^{13}$CO)/($dv/dr$) of
$1\times10^{-6}$~(km~s$^{-1}$~pc$^{-1}$)$^{-1}$ can be explained as
the Galactic abundances of [$^{13}$CO]/[H$_{2}$] = $1\times10^{-6}$
\citep{sol79} with the velocity gradient of 1~km~s$^{-1}$~pc$^{-1}$.
Other parameters are the same as in Sect.~\ref{dis-co}.

\begin{figure}
\plotone{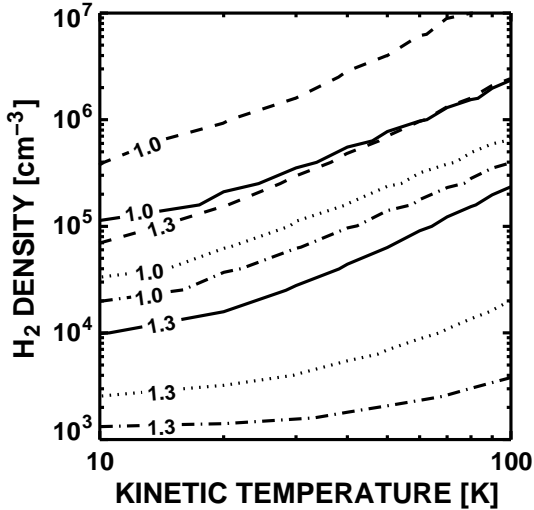}
\caption{The LVG calculation results for the
	$^{13}$CO(2-1)/C$^{18}$O(2-1) ratio as a function of H$_{2}$
	number density and kinetic temperature.
	Solid, dashed, and dotted lines are the
	$^{13}$CO(2-1)/C$^{18}$O(2-1) ratios with
	$Z$($^{13}$CO)/($dv/dr$) of $1\times10^{-6}$, $1\times10^{-7}$,
	and $1\times10^{-5}$~(km~s$^{-1}$~pc$^{-1}$)$^{-1}$,
	respectively, under the [$^{13}$CO]/[C$^{18}$O] relative
	abundance ratio of 4.
	Dot-dashed lines are the $^{13}$CO(2-1)/C$^{18}$O(2-1) ratios
	with the [$^{13}$CO]/[C$^{18}$O] relative abundance ratio of 2
	under $Z$($^{13}$CO)/($dv/dr$) of
	$1\times10^{-6}$~(km~s$^{-1}$~pc$^{-1}$)$^{-1}$.
\label{fig-lvg}}
\end{figure}

The calculation results are shown in Fig.~\ref{fig-lvg}.
In the case of $Z$($^{13}$CO)/($dv/dr$) of
$1\times10^{-6}$~(km~s$^{-1}$~pc$^{-1}$)$^{-1}$, a high density of
$n_{\rm H_{2}} > 1 \times 10^{4}$~cm$^{-3}$ is needed even for
$T_{\rm k}$ of 10~K, and about an order higher density is needed for
100~K to realize the $^{13}$CO(2-1)/C$^{18}$O(2-1) ratio of
$1.0\pm0.3$.
This is because both the $^{13}$CO and C$^{18}$O lines easily become
optically thin at lower-J with the increase of temperature, since the
population moves to higher-J.
To compensate this, the density, and therefore the column density per
unit velocity,
$N$($^{13}$CO or C$^{18}$O)/$dv = Z$($^{13}$CO or C$^{18}$O)/($dv/dr$)
$\times n_{\rm H_{2}}$, also has to be high for both lines to be
optically thick.
If we increase $Z$($^{13}$CO)/($dv/dr$) by an order of magnitude, the
density decreases by about a factor of several at a certain
temperature.
This is because the increase of $Z$($^{13}$CO)/($dv/dr$) makes the
line easier to be optically thick.
The response will be opposite if we decrease $Z$($^{13}$CO)/($dv/dr$)
by an order of magnitude.

In the case of a lower [$^{13}$CO]/[C$^{18}$O] relative abundance
ratio of 2 (half the abundance ratio used above with increasing
[C$^{18}$O]), the required density for the ratio of $1.0\pm0.3$ is
about an order of magnitude lower at a certain temperature
(Fig.~\ref{fig-lvg}).
This is because the C$^{18}$O abundance increased from the above
condition, the opacity and therefore the intensity of the C$^{18}$O
line become similar to that of the $^{13}$CO line.

As shown above, the important parameters for the low
$^{13}$CO(2-1)/C$^{18}$O(2-1) ratio are (1) the molecular gas density
and hence the column density per unit velocity, and (2) the molecular
abundance.
First, we discuss the gas density.
The molecular gas density needs to be high
($\gtrsim10^{4}$~cm$^{-3}$) for the ratio to be around unity.
The molecular gas in Arp 220 indeed contains high density gas, which
is supported by the observations of high critical density molecules,
such as HCN, HCO$^{+}$ or CS \citep[e.g.,][]{sol90,sol92,gre08}.
On the other hand, there are many galaxies with the detections of
these high critical density molecular lines, but almost no report of
a $^{13}$CO(2-1)/C$^{18}$O(2-1) $\sim1$ so far.
One possibility of the difference between Arp 220 and other galaxies
may be due to the large fraction of dense molecular gas.
Our $^{13}$CO(2-1) and C$^{18}$O(2-1) images exhibit compact
distribution around the double nucleus, and the molecular gas in
Arp 220 is dominated by dense gas \citep[e.g.,][]{gre08}.
If the dense gas is concentrated toward the double nucleus, most of
the molecular gas toward the double nucleus can be dominated by dense
gas, and this makes the column density high enough to result in the
$^{13}$CO(2-1)/C$^{18}$O(2-1) of unity.

Second, we discuss the molecular abundance.
To realize the $^{13}$CO(2-1)/C$^{18}$O(2-1) intensity ratio of about
unity with changing the abundance, two possibilities can be
considered; one is the underabundance of the $^{13}$CO molecule, and
another is the overabundance of C$^{18}$O.
Deficiency of $^{13}$CO abundance is often suggested in merging
galaxies based on their larger $^{12}$CO(2-1)/$^{13}$CO(2-1) ratios
than those in starburst or Seyfert galaxies
\citep[e.g.,][]{aal91,cas92}.
But as is mentioned in Sect.~\ref{res-ratio}, the observed
$^{12}$CO(2-1)/$^{13}$CO(2-1) ratio of Arp 220 is not as extreme as
those in the other merging galaxies, and not significantly different
from those in starburst or Seyfert galaxies.
In addition, the possible reason for the $^{13}$CO deficiency is the
selective photodissociation of the $^{13}$CO molecules
\citep[e.g.,][]{cas92}.
In this case, however, C$^{18}$O molecules will be more affected
by the selective photodissociation \citep{dis88,cas92}, hence this
cannot be the cause.
We therefore think that the underabundance of $^{13}$CO is possible,
but less likely.

The overabundance of the C$^{18}$O molecule may be possible to
achieve under the circumstance of Arp 220.
Massive stars synthesize a large amount of the primary element,
$^{12}$C, at helium burning phase, and it goes into interstellar
medium via supernova explosions \citep{cas92}.
The $^{18}$O enrichment occurs also in massive stars
\citep{hen93,sag91}, either Wolf-Rayet stars or type II supernova
explosions by partial helium burning \citep{ama95}.
Since Arp 220 is very active in star formation (Sect.~\ref{intro}),
both the $^{12}$C and $^{18}$O enrichment due to the above mechanism
can be realized.
This can lead to the enrichment of the C$^{18}$O molecule.
This possibility still needs to be studied quantitatively.
Note that a recent molecular gas abundance study toward a young
(several Gyr old) galaxy at $z=0.89$ show low [$^{13}$CO]/[C$^{18}$O]
of $1.9\pm0.2$ \citep{mul06}.
This result also suggests that the low intensity ratio is related to
an abundance anomaly during the young star formation epoch.

\section{CONCLUSIONS}
\label{concl}

We observed the central region of Arp 220 in the $^{12}$CO(6-5),
$^{12}$CO(2-1), $^{13}$CO(2-1), and C$^{18}$O(2-1) lines, and
435~$\mu$m and 1.3~mm continuum simultaneously with the SMA.
The two nuclei are clearly resolved in the 435~$\mu$m image, and
kinematically resolved in the $^{12}$CO(6-5) image.

For the double nucleus, we concluded as follows:
\begin{itemize}
\item The difference of the peak intensities in our 435~$\mu$m image
	between the two nuclei is smaller than at longer wavelengths.
	From the dust SED fitting, the dust in the two nuclei is
	estimated to be optically thick at 435~$\mu$m.
	The emissivity indices are estimated to be $\sim2.0$ for the
	eastern nucleus and $\sim1.0$ for the western nucleus.
	This suggests that the dust properties, such as dust size
	distributions or dust compositions, are different between the two
	nuclei.
\item The $^{12}$CO SEDs are similar between the two nuclei with the
	western nucleus having higher upper limits in the excitation
	conditions than those in the eastern nucleus.
	The $^{12}$CO SEDs for both nuclei and that of M82 or
	BR 1202--0725 are similar, characterized with small intensity
	decreases up to J = 6-5 ($^{12}$CO(6-5)/(2-1) ratio of about
	0.5).
	This suggests that the molecular gas in the two nuclei of
	Arp 220 has the similar excitation conditions as that in M82 or
	BR 1202--0725, which have a density of
	$\gtrsim10^{3.3}$~cm$^{-3}$ or a temperature of $\gtrsim30$~K.
\item We could not find any evidence of an AGN in Arp 220 with the
	$^{12}$CO SED study.
	There is no clear difference in the $^{12}$CO SEDs between the
	AGN hosting ULIRG Mrk 231 and the double nucleus of Arp 220 (and
	therefore M82 and BR 1202--0725).
	This suggests that the AGN heating is not important for molecular
	gas excitation conditions in the large scale (a few hundred to a
	few kpc scale).
\end{itemize}
For the global characteristics of the molecular gas and dust in
Arp 220, we concluded as follows:
\begin{itemize}
\item Based on the large amount of missing flux in our data and other
	previously published evidence in molecular gas and dust, we
	suggest the existence of an extended component in the dust
	emission with its dust properties $T_{\rm d}\sim38$~K,
	$\beta\sim2.4$, and $\nu_{\rm c}\sim200~\micron$.
	A recent SMA U/LIRG survey suggests that many of U/LIRGs seem to
	have extended components \citep{wil08}, so that having an
	extended dust component might be common.
\item The $^{12}$CO(2-1) line spectrum shows stronger line intensity
	at the lower velocities than the higher velocities, but the
	spectra of the higher-J lines show the opposite, indicating that
	the higher velocity gas has higher density, higher temperature,
	or both, than the lower velocity component.
\item The intensities of the $^{13}$CO(2-1) and C$^{18}$O(2-1) lines
	are similar.
	This suggests that the molecular gas in Arp 220 is dense
	enough to be optically thick in both lines, or the abundance of
	either line deviates from the values in other nearby galaxies.
	To explain the ratio with the density effect, Arp 220 should have
	molecular gas largely dominated by dense gas, more than in other
	nearby galaxies.
	Underabundance of $^{13}$CO is possible, but it is less likely.
	Overabundance of C$^{18}$O is also possible, considering the
	$^{12}$C and $^{18}$O enrichment by high mass stars.
\end{itemize}

\acknowledgements

We thank all the past and present SMA staff for designing,
constructing, and supporting the SMA, especially who worked for and
realized the 690~GHz observations.
We also thank the anonymous referee for helpful comments.
The Submillimeter Array is a joint project between the Smithsonian
Astrophysical Observatory and the Academia Sinica Institute of
Astronomy and Astrophysics and is funded by the Smithsonian
Institution and the Academia Sinica.
This work is supported by the National Science Council (NSC) of
Taiwan, NSC 97-2112-M-001-021-MY3.


\begin{thebibliography}{99}
\bibitem[Aalto et al.(1991)]{aal91} Aalto, S., Black, J. H.,
	Johansson, L. E. B., \& Booth, R. S.  1991, \aap, 249, 323
\bibitem[Aalto et al.(1995)]{aal95} Aalto, S., Booth, R. S.,
	Black, J. H., \& Johansson, L. E. B.  1995, \aap, 300, 369
\bibitem[Amari et al.(1995)Amari, Zinner, \& Lewis]{ama95} Amari, S.,
	Zinner, E., \& Lewis, R. S.  1995, \apjl, 447, L147
\bibitem[Arp(1966)]{arp66} Arp, H.  1966, \apjs, 14, 1
\bibitem[Barnes \& Hernquist(1991)]{bar91} Barnes, J. E.,
	\& Hernquist, L. E.  1991, \apjl, 370, L65
\bibitem[Baudry \& Neri(2000)]{bau00} Baudry, A., \& Neri, R.  2000,
	in Proceeding from IRAM Millimeter Interferometry Summer School
	2, ed. A. Dutrey, 233
\bibitem[Borys et al.(2003)]{bor03} Borys, C., Chapman, S.,
	Halpern, M., \& Scott, D.  2003, \mnras, 344, 385
\bibitem[Carico et al.(1992)]{car92} Carico, D. P., Keene, J.,
	Soifer, B. T., \& Neugebauer, G.  1992, \pasp, 104, 1086
\bibitem[Carilli et al.(2002)]{car02} Carilli, C. L., Kohno, K.,
	Kawabe, R., Ohta, K., Henkel, C., Menten, K. M., Yun, M. S.,
	Petric, A., \& Tutui, Y.  2002, \aj, 123, 1838
\bibitem[Casoli et al.(1992)]{cas92} Casoli, F., Dupraz, C.,
	\& Combes, F.  1992, \aap, 264, 55
\bibitem[Chou et al.(2007)]{cho07} Chou, R. C.-Y., Peck, A. B.,
	Lim, J., Matsushita, S., Muller, S., Sawada-Satoh, S.,
	Dinh-V-Trung, Boone, F., \& Henkel, C. 2007, \apj, 670, 116
\bibitem[Downes \& Eckart(2007)]{dow07} Downes, D.,
	\& Eckart, A.  2007, \aap, 468, L57
\bibitem[Downes \& Solomon(1998)]{dow98} Downes, D.,
	\& Solomon, P. M.  1998, \apj, 507, 615
\bibitem[Dunne \& Eales(2001)]{dun01} Dunne, L.,
	\& Eales, S. A.  2001, \mnras, 327, 697
\bibitem[Eales et al.(1999)]{eal99} Eales, S., Lilly, S., Gear, W.,
	Dunne, L., Bond, J. R., Hammer, F., Le Fe\`vre, O.,
	\& Crampton, D.  1999, \apj, 515, 518
\bibitem[Eales et al.(2000)]{eal00} Eales, S., Lilly, S., Webb, T.,
	Dunne, L., Gear, W., Clements, D., \& Yun, M.  2000, \aj, 120,
	2244
\bibitem[Eales et al.(1989)Eales, Wynn-Williams, \& Duncan]{eal89}
	Eales, S. A., Wynn-Williams, C. G., \& Duncan, W. D.  1989,
	\apj, 339, 859
\bibitem[Fixsen et al.(1999)Fixsen, Bennet, \& Mather]{fix99}
	Fixsen, D. J., Bennett, C. L., \& Mather, J. C.  1999, \apj, 526,
	207
\bibitem[Flower \& Launay(1985)]{flo85} Flower, D. R.,
	\& Launay, J. M.  1985, \mnras, 214, 271
\bibitem[Gao \& Solomon(2004)]{gao04} Gao, Y.,
	\& Solomon, P. M.  2004, \apj, 606, 271
\bibitem[Glenn \& Hunter(2001)]{gle01} Glenn, J.,
	\& Hunter, T. R.  2001, \apjs, 135, 177
\bibitem[Goldreich \& Kwan(1974)]{gol74} Goldreich, P.,
	\& Kwan, J.  1974, \apj, 189, 441
\bibitem[Gonz\'alez-Alfonso et al.(2004)]{gon04}
	Gonz\'alez-Alfonso, E., Smith, H. A., Fischer, J.,
	\& Cernicharo, J.  2004, \apj, 613, 247
\bibitem[Graham et al.(1990)]{gra90} Graham, J. R., Carico, D. P.,
	Matthews, K., Neugebauer, G., Soifer, B. T.,
	\& Wilson, T. D.  1990 \apjl, 354, L5
\bibitem[Greve et al.(2005)]{gre05} Greve, T. R., et al.  2005,
	\mnras, 359, 1165
\bibitem[Greve et al.(2008)]{gre08} Greve, T. R.,
	Papadopoulos, P. P., Gao, Y., \& Radford, S. J. E.  2008,
	\apj, submitted (astro-ph/0610378)
\bibitem[Henkel \& Mauersberger(1993)]{hen93} Henkel, C.,
	\& Mauersberger, R.  1993, \aap, 274, 730
\bibitem[Hernquist(1992)]{her92} Hernquist, L.  1992, \apj, 400, 460
\bibitem[Hernquist(1993)]{her93} Hernquist, L.  1993, \apj, 409, 548
\bibitem[Ho et al.(2004)]{ho04} Ho, P. T. P., Moran, J. M.,
	\& Lo, F.  2004, \apjl, 616, L1
\bibitem[Hsieh et al.(2008)]{hsi08} Hsieh, P.-Y., Matsushita, S.,
	Lim, J., Kohno, K., \& Sawada-Satoh, S.  2008, \apj, 683, 70
\bibitem[Hughes et al.(1998)]{hug98} Hughes, D. H., et al.  1998,
	\nat, 394, 241
\bibitem[Joseph \& Wright(1985)]{jos85} Joseph, R. D.,
	\& Wright, G. S.  1985, \mnras, 214, 87
\bibitem[Klass et al.(2001)]{kla01} Klaas, U., et al.  2001, \aap,
	379, 823
\bibitem[Lonsdale et al.(2006)]{lon06} Lonsdale, C. J.,
	Diamond, P. J., Thrall, H., Smith, H. E.,
	\& Lonsdale, C. J.  2006, \apj, 647, 185
\bibitem[Matsushita et al.(2004)]{mat04}
    Matsushita, S., et al.  2004, \apjl, 616, L55
\bibitem[McKee et al.(1982)]{mck82} McKee, C. F., Storey, J. W. V.,
	Watson, D. W., \& Green, S.  1982, \apj, 259, 647
\bibitem[Meier \& Turner(2004)]{mei04} Meier, D. S.,
	\& Turner, J. L.  2004, \aj, 127, 2069
\bibitem[Mihos \& Hernquist(1996)]{mih96} Mihos, J. C.,
	\& Hernquist, L.  1996, \apj, 464, 641
\bibitem[Muller et al.(2006)]{mul06} Muller, S., Gu\'elin, M.,
	Dumke, M., Lucas, R., \& Combes, F.  2006, \aap, 458, 417
\bibitem[Norris(1988)]{nor88} Norris, R. P.  1988, \mnras, 230, 345
\bibitem[Ohta et al.(1996)]{oht96} Ohta, K., Yamada, T.,
	Nakanishi, K., Kohno, K., Akiyama, M., \& Kawabe, R.  1996, \nat,
	382, 426
\bibitem[Omont et al.(1996)]{omo96} Omont, A., Petitjean, P.,
	Guilloteau, S., McMahon, R. G., Solomon, P. M.,
	\& P\'econtal, E.  1996, \nat, 382, 428
\bibitem[Papadopoulos \& Seaquist(1998)]{pap98} Papadopoulos, P. P.,
	\& Seaquist, E. R.  1998, \apj, 492, 521
\bibitem[Papadopoulos et al.(2007)]{pap07} Papadopoulos, P. P.,
	Isaak, K. G., \& van der Werf, P. P.  2007, \apj, 668, 815
\bibitem[Parra et al.(2007)]{par07} Parra, R., Conway, J. E.,
	Diamond, P. J., Thrall, H., Lonsdale, C. J., Lonsdale, C. J.,
	\& Smith, H. E.  2007, \apj, 659, 314
\bibitem[Petitpas \& Wilson(2000)]{pet00} Petitpas, G. R.,
	\& Wilson, C. D.  2000, \apjl, 538, L117
\bibitem[Riechers et al.(2006)]{rie06} Riechers, D. A., et al.  2006,
	\apj, 650, 604
\bibitem[Rigopoulou et al.(1996)]{rig96} Rigopoulou, D.,
	Lawrence, A., \& Rowan-Robinson, M.  1996, \mnras, 278, 1049
\bibitem[Sage et al.(1991)]{sag91} Sage, L. J., Mauersberger, R.,
	\& Henkel, C.  1991, \aap, 249, 31
\bibitem[Sakamoto et al.(2006a)]{sak06a} Sakamoto, K., et al.  2006,
	\apj, 636, 685
\bibitem[Sakamoto et al.(2007)]{sak07} Sakamoto, K., Ho, P. T. P.,
	Mao, R.-Q., Matsushita, S., \& Peck, A. B.,  2007, \apj, 654, 782
\bibitem[Sakamoto et al.(2006b)Sakamoto, Ho, \& Peck]{sak06b}
	Sakamoto, K., Ho, P. T. P. \& Peck, A. B.  2006, \apj, 644, 862
\bibitem[Sakamoto et al.(1999)]{sak99} Sakamoto, K., Scoville, N. Z.,
	Yun, M. S., Crosas, M., Genzel, R., \& Tacconi, L. J.  1999,
	\apj, 514, 68
\bibitem[Sakamoto et al.(2008)]{sak08} Sakamoto, K., Wang, J.,
	Wiedner, M. C., Wang, Z., Peck, A. B., Zhang, Q.,
	Petitpas, G. R., Ho, P. T. P., \& Wilner, D.  2008, \apj, 684,
	957
\bibitem[Sanders et al.(2003)]{san03} Sanders, D. B.,
	Mazzarella, J. M., Kim, D.-C., Surace, J. A.,
	\& Soifer, B. T.  2003, \aj, 126, 1607
\bibitem[Scott et al.(2002)]{sco02} Scott, S. E., et al.  2002,
	\mnras, 331, 817
\bibitem[Scoville et al.(1998)]{sco98} Scoville, N. Z., et al.  1998,
	\apjl, 492, L107
\bibitem[Scoville et al.(1991)]{sco91} Scoville, N. Z.,
	Sargent, A. I., Sanders, D. B., \& Soifer, B. T.  1991, \apjl,
	366, L5
\bibitem[Scoville et al.(1997)Scoville, Yun, \& Bryant]{sco97}
	Scoville, N. Z., Yun, M. S., \& Bryant, P. M.  1997, \apj, 484,
	702
\bibitem[Smith et al.(1998)]{smi98} Smith, H. E., Lonsdale, C. J.,
	Lonsdale, C. J., \& Diamond, P. J.  1998, \apjl, 493, L17
\bibitem[Soifer et al.(1999)]{soi99} Soifer, B. T., Neugebauer, G.,
	Matthews, K., Becklin, E. E., Ressler, M., Werner, M. W.,
	Weinberger, A. J., \& Egami, E.  1999, \apj, 513, 207
\bibitem[Solomon et al.(1992)]{sol92} Solomon, P. M., Downes, D.,
	\& Radford, S. J. E.  1992, \apj, 387, L55
\bibitem[Solomon et al.(1990)]{sol90} Solomon, P. M.,
	Radford, S. J. E., \& Downes, D.  1990, \apjl, 348, L53
\bibitem[Solomon et al.(1979)]{sol79} Solomon, P. M.,
	Scoville, N. Z., \& Sanders, D. B.  1979, \apjl, 232, L89
\bibitem[Tacconi et al.(2006)]{tac06} Tacconi, L. J., et al. 2006,
	\apj, 640, 228
\bibitem[van Dishoeck \& Black(1988)]{dis88} van Dishoeck, E. F.,
	\& Black, J. H.  1988, \apj, 334, 771
\bibitem[Wang et al.(2004)]{wan04} Wang, M., Henkel, C., Chin, Y.-N.,
	Whiteoak, J. B., Cunningham, M. H., Mauersberger, R.,
	\& Muders, D.  2004, \aap, 422, 883
\bibitem[Wei\ss\ et al.(2005a)]{wei05a} Wei\ss, A., Walter, F.,
	\& Scoville, N. Z.  2005a, \aap, 438, 533
\bibitem[Wei\ss\ et al.(2005b)]{wei05b} Wei\ss, A., Downes, D.,
	Walter, F., \& Henkel, C.  2005b, \aap, 440, L45
\bibitem[Wei\ss\ et al.(2007)]{wei07} Wei\ss, A., Downes, D.,
	Neri, R., Walter, F., Henkel, C., Wilner, D. J., Wagg, J.,
	\& Wiklind, T.  2007, \aap, 467, 955
\bibitem[Wiedner et al.(2002)]{wie02} Wiedner, M. C., Wilson, C. D.,
	Harrison, A., Hills, R. E., Lay, O. P.,
	\& Carlstron, J. E.  2002, \apj, 581, 229
\bibitem[Wilson et al.(2008)]{wil08} Wilson, C. D., et al.  2008,
	\apjs, 178, 189
\bibitem[Woody et al.(1989)]{woo89} Woody, D. P., Scott, S. L.,
	Scoville, N. Z., Mundy, L. G., Sargent, A. I., Padin, S.,
	Tinney, C. G., \& Wilson, C. D.  1989, \apjl, 337, L41
\end{thebibliography}
\end{document}